\documentclass[aps,pre,onecolumn,floatfix]{revtex4}

\usepackage{bm}
\usepackage{graphicx}
\usepackage{amssymb,amsfonts,amsmath}
\usepackage{epsf}
\usepackage{color}
\usepackage{hyperref}
\usepackage{subfigure}
\usepackage[export]{adjustbox}
\usepackage{epstopdf}
\usepackage{rotating}
\usepackage{soul}

\usepackage{mathrsfs}

\usepackage{upgreek}

\usepackage[T3,T1]{fontenc}
\DeclareSymbolFont{tipa}{T3}{cmr}{m}{n}
\DeclareMathAccent{\invbreve}{\mathalpha}{tipa}{16}

\newcommand{\be}{\begin{equation}}
\newcommand{\ee}{\end{equation}}
\newcommand{\bea}{\begin{eqnarray}}
\newcommand{\eea}{\end{eqnarray}}

\begin{document}

\title{\bf Dynamical response and time correlation functions in random quantum systems}

\author{Sudhir Ranjan Jain}
\email{srjain@cbs.ac.in; \vfill\break ORCID: 0000-0002-0472-5549}
\affiliation{UM-DAE Centre for Excellence in Basic Sciences, University of Mumbai, Vidyanagari Campus, Mumbai 400098, India}
\author{Pierre Gaspard}
\email{Gaspard.Pierre@ulb.be; \vfill\break ORCID: 0000-0003-3804-2110}
\affiliation{Center for Nonlinear Phenomena and Complex Systems, Universit{\'e} Libre de Bruxelles (U.L.B.), Code Postal 231, Campus Plaine, B-1050 Brussels, Belgium}

\begin{abstract}
{\it Abstract:} Time-dependent response and correlation functions are studied in random quantum systems composed of infinitely many parts without mutual interaction and defined with statistically independent random matrices. The latter are taken within the three Wigner-Dyson universality classes.  In these systems, the response functions are shown to be exactly given by statistical averages over the random-matrix ensemble.  Analytical results are obtained for the time dependence of the mean response and correlation functions at zero and positive temperatures. At long times, the mean correlation functions are shown to have a power-law decay for GOE at positive temperatures, but for GUE and GSE at zero temperature.  Otherwise, the decay is much faster in time.  In relation to these power-law decays, the associated spectral densities have a dip around zero frequency.  The diagrammatic method is developed to obtain higher-order response functions and the third-order response function is explicitly calculated.  The response to impulsive perturbations is also considered.  In addition, the quantum fluctuations of the correlation function in individual members of the ensemble are characterised in terms of their probability distribution, which is shown to change with the temperature.

\vskip 0.2 cm


\end{abstract}

\maketitle

%%%%%%%%%%%%%%%%%%%%%%%%%%%%%%%%%%%%%%%%%%%%%%%%%
\section{Introduction}
\label{sec:intro}

In physical sciences, a fundamental result is that rate and transport coefficients can be calculated from the microscopic dynamics in terms of time-dependent response and correlation functions.  In particular, time-dependent dipole-dipole autocorrelation functions determine leading photoabsorption cross-sections in nuclei, atoms, molecules, and nanoparticles \cite{M95,GAB95,GJ97,T07}.  In nuclear physics, the concept of nuclear friction is also related to time correlation functions \cite{BSS93,H97}.  In addition, at the macroscale, transport coefficients for viscosity, heat conduction, and diffusion can be computed using Green-Kubo formulas involving current-current autocorrelation functions \cite{G52,G54,K57,MG23}.  These developments can be justified on the basis of time-dependent perturbation theory, which leads to linear response theory \cite{CW51} and nonlinear response generalisations \cite{BC59}, especially, for nonlinear optical spectroscopy \cite{M95}. The time-dependent approach is complementary to the time-independent one, both being related to each other by Fourier or Laplace transforms \cite{T07}.

Accordingly, the response properties of any system are given by response or correlation functions of the observables coupling the system to external fields and evolving in time under the internal dynamics of the system itself.  The response and correlation functions should be evaluated in the pure or thermal state of the unperturbed system before the arrival of the external perturbation.  From few-body to many-body quantum systems, the response and correlation functions are thus largely determined on the Hamiltonian operator and its spectral properties.  The system may have symmetries, leading to the decomposition of the Hamiltonian into blocks corresponding to the different symmetry sectors of the quantum state space.  Within each symmetry sector, the generic behaviour is that the energy eigenvalues are subjected to Wigner level repulsion, inducing random-matrix effects on small energy scales \cite{W59}.  These effects fall into the Wigner-Dyson universality classes corresponding to the orthogonal, unitary, and symplectic random-matrix ensembles \cite{W59,D62a,D62b,D62c,D62d,M67,H01}.  However, the large energy scales are described by the density of states (DoS), which is always specific to the system of interest.  If the quantum system is bounded and finite, the energy spectrum is discrete, but the DoS increases very quickly (exponentially) with the number of degrees of freedom, so that the spectrum is soon quasi continuous.  In the time domain, the universal features of the energy spectrum are thus expected to manifest themselves on long time scales. In this context, a key issue is to understand the relative magnitude of the aforementioned universal features with respect to the specific properties associated with the DoS of the system and the particular observables of coupling to external fields.

In this paper, our purpose is to investigate this issue for special kinds of random quantum systems composed of infinitely many parts without mutual interaction, each part having a Hamiltonian and a coupling operator given by random Hermitian matrices.  Examples of such systems are given by well separated subsystems like nuclear spins or impurities embedded in a disordered environment, like two-level subsystems in glasses \cite{SS94}.  Random Hermitian matrices have also been considered to model Hamiltonians of heat reservoirs, to which a spin is coupled \cite{EG03}.  The literature contains studies of different properties for such systems.  On the one hand, the correlation between the eigenvalues of general random Hermitian matrices have been reported on in Refs.~\cite{BZ93,BZ94}.  On the other hand, the time-dependent correlation functions of random-matrix operators have been studied in Refs.~\cite{AF00a,AF00b}, but only in the infinite-temperature limit without envisaging the response functions in thermal states.  The issue is also of interest in the study of classically chaotic quantum systems, which are known to manifest Wigner level repulsion on small energy scales \cite{H01,BGS84,HMABH07}.

The paper is organized as follows.  In Sec.~\ref{sec:systems}, we define the random quantum systems we consider on the basis of the orthogonal, unitary, and symplectic random-matrix ensembles.  The response theory of these systems is presented in Sec.~\ref{sec:response}.  Section~\ref{sec:1st-resp} is devoted to the mean first-order response function and the related mean two-time correlation function at early, intermediate, and long times and different temperatures for the three Wigner-Dyson universality classes.  The behaviours of the associated spectral density are also considered.  In Sec.~\ref{sec:higher-resp}, the mean third-order response function and the related mean four-time correlation function are calculated at intermediate times and a diagrammatic method is developed to obtain still higher-order response and correlation functions.  The contributing diagrams are shown to be enumerated in terms of polynomials related to, but different from the Bell polynomials.  These results are applied to the response to impulsive perturbations in Sec.~\ref{sec:impulsive}.  In Sec.~\ref{sec:fluctuations}, we characterise the quantum fluctuations of individual members of the ensemble in terms of their probability distribution in different regimes depending on the temperature.  Conclusion and perspectives are drawn in Sec.~\ref{sec:conclusion}.
\\

{\it Notations.} The trace of some matrix or operator is denoted $``{\rm tr}"$ and is supposed to apply on the whole expression to the right-hand side of this symbol. $k_{\rm B}$ denotes Boltzmann's constant and $T$ the temperature.

%%%%%%%%%%%%%%%%%%%%%%%%%%%%%%%%%%%%%%%%%%%%%%%%%
\section{Random quantum systems}
\label{sec:systems}

\subsection{Definition}

We consider quantum systems composed of many parts such as nuclear spins, atoms, molecules, or point defects that are embedded in a disordered environment like a glassy material, where their interactions are negligible except for thermalization at the temperature of the environment.  Accordingly, the quantum system is supposed to have $P$~parts without mutual interaction and it may be subjected to some external field $\lambda$.  The total Hamiltonian ruling the system is given by
\be
\hat H^{(P)} = \hat H_0^{(P)} + \lambda \, \hat V^{(P)} \, ,
\ee
where the unperturbed Hamiltonian and the perturbation operator have the following forms,
\bea
&& \hat H_0^{(P)}  = \sum_{p=1}^P \hat H_{0p} \equiv \sum_{p=1}^{P} \hat I \otimes \cdots \otimes \hat I \otimes \underbrace{\hat H_{0p}}_{p^{\rm th}\, {\rm factor}} \otimes \hat I \otimes \cdots \otimes \hat I \, , \label{dfn-H0-P}\\
&& \hat V^{(P)}  = \sum_{p=1}^P \hat V_{p} \equiv \sum_{p=1}^{P} \hat I \otimes \cdots \otimes \hat I \otimes \underbrace{\hat V_{p}}_{p^{\rm th}\, {\rm factor}} \otimes \hat I \otimes \cdots \otimes \hat I \, , \label{dfn-V-P}
\eea
$\hat H_{0p}$ and $\hat V_p$ being Hermitian matrices and $\hat I$ denoting the corresponding identity matrix, all of dimensions $D\times D$.  Thus, the state space of the total system ${\cal E}^{(P)} = {\cal E}_1\otimes{\cal E}_2\otimes\cdots\otimes{\cal E}_P$ has the dimension ${\rm dim}\,{\cal E}^{(P)} = \prod_{p=1}^P  {\rm dim}\,{\cal E}_p = D^P$.  The form of the Hamiltonian expresses the assumption that there is no interaction between the parts of the system.  

We assume that each one of the matrices $\hat H_{0p}$ and $\hat V_p$ for $p=1,2,\dots,P$ is a random Hermitian matrix taken within a particular statistical ensemble defined with some statistically independent probability distributions ${\cal P}(\hat H_0)$ and ${\cal P}(\hat V)$.  For instance, we may consider the Gaussian ensembles
\bea
{\cal P}(\hat H_0) &=& C_{N\nu a_{H_0}} \, \exp\left(-\frac{\nu}{2} \, a_{H_0} \, {\rm tr}\, \hat H_0^2 \right) , \label{P(H0)}\\
{\cal P}(\hat V) &=& C_{N\nu a_{V}} \, \exp\left(-\frac{\nu}{2} \, a_{V} \, {\rm tr}\, \hat V^2 \right) , \label{P(V)}
\eea
where $C_{N\nu a}$ are the corresponding normalisation constants and $\nu=1,2,4$ for respectively, orthogonal (GOE), unitary (GUE), and symplectic (GSE) ensembles \cite{W59,D62a,D62b,D62c,D62d,M67,H01}.  Since $\hat H_0$ and $\hat V$ are assumed to be statistically independent, the perturbed Hamiltonian $\hat H=\hat H_0+\lambda \hat V$ has a probability distribution similar to Eq.~(\ref{P(H0)}) but with a parameter given by
\be
\frac{1}{a_H} = \frac{1}{a_{H_0}} + \frac{\lambda^2}{a_V} \, .
\ee
Statistical averages over matrix ensembles defined by the probability distributions~(\ref{P(H0)}) and~(\ref{P(V)}) are denoted $\langle\cdot\rangle_{\rm ME}$.

\subsection{The random-matrix ensembles and their basic properties}

Here, the orthogonal, unitary, and symplectic matrices are defined and the main properties of the corresponding random-matrix ensembles are presented \cite{W59,D62a,D62b,D62c,D62d,M67,H01}.

In the GOE, the $N\times N$ matrices are real symmetric composed of $\frac{1}{2}N(N+1)$ real numbers.  

In the GUE, the $N\times N$ matrices are complex Hermitian containing $N^2$ real numbers. 

In the GSE, the Hamiltonian operators are $2N\times 2N$ matrices of the following form,
\be\label{symplectic}
\hat H =
\left(\begin{array}{cc}
\hat H_0 -{\rm i}\, \hat H_3 & -{\rm i}\, \hat H_1 - \hat H_2 \\
-{\rm i}\, \hat H_1 + \hat H_2 & \hat H_0 +{\rm i}\, \hat H_3 \\
\end{array}\right) ,
\ee
where $\hat H_i$ with $i=0,1,2,3$ are $N\times N$ real matrices such that their transpose satisfies $\hat H_0^{\rm T}=\hat H_0$ and $\hat H_j^{\rm T}=-\hat H_j$ for $j=1,2,3$.  The matrices~(\ref{symplectic}) are not only Hermitian $\hat H=\hat H^{\dagger}$, but also invariant under the time-reversal transformation $\hat\Theta = \hat\Sigma\hat K$, where $\hat\Sigma=\left(\begin{array}{cc}
0 & - \hat I_N \\
\hat I_N & 0 \\
\end{array}\right)$ and $\hat K$ is the operation of complex conjugate, i.e., $\hat H=\hat\Theta\hat H \hat\Theta^{-1}$.  Therefore, their $2N$ eigenvalues are real and they have the multiplicity two, which is called Kramers' degeneracy.

The average density of states over some matrix ensemble is defined as $\sigma(E)=\langle\invbreve{\sigma}(E)\rangle_{\rm ME}$ with $\invbreve{\sigma}(E)={\rm tr}\, \delta(E-\hat H)=\sum_k\delta(E-E_k)$, where the sum extends over the eigenstates of the Hamiltonian matrix $\hat H$, $\{E_k\}$ being the associated energy eigenvalues and $\delta(x)$ denoting the Dirac delta distribution.  We note that the eigenstates should be written in general as $\vert k\rangle=\vert m,i\rangle$, where $m$ is the label of the eigenvalue $E_k=E_m$ and $i\in\{1,g_{\nu}\}$ is the index of multiplicity of the eigenvalue $E_m$.  The state space has thus the dimension $D={\rm dim}\,{\cal E}=g_\nu N$, while the number of eigenvalues is equal to $N$.

For large $N$, the average density of states is given by the Wigner semicircle law \cite{W59},
\be\label{semicircle_law}
\sigma(E) = \left\{
\begin{array}{ll}
\displaystyle g_{\nu} \, \frac{a_H}{\pi}\, \sqrt{\frac{2N}{a_H}- E^2} &\qquad\mbox{for} \quad \vert E\vert < \sqrt{\frac{2N}{a_H}} \, , \\
0  &\qquad\mbox{otherwise} \, , \\
\end{array}\right.
\ee
with the multiplicities $g_1=g_2=1$ and $g_4=2$.  Therefore, the energy spectrum forms a quasi-continuous energy band, extending over the interval $-E_{\rm b} < E < +E_{\rm b}$ with the band half-width $E_{\rm b}=\sqrt{\frac{2N}{a_H}}$.

Furthermore, the variances of the matrix elements of the perturbation operator are given by
\be\label{var-V}
\langle\vert V_{kl}\vert^2\rangle_{\rm ME} = \frac{1}{2 a_V} \left( 1 + \gamma_{\nu} \, \delta_{kl}\right)
\qquad\mbox{with}\qquad
\gamma_{\nu} = \frac{2}{\nu}-1 \, ,
\ee
so that $\gamma_1=1$, $\gamma_2=0$, and $\gamma_4=-\frac{1}{2}$, as shown in Appendix~\ref{app:RMT}.

On small energy scales, the eigenvalue spectra of random Hermitian matrices obey universal statistical properties.  In particular, the nearest-neighbour level-spacing distribution behaves as
\be\label{P(s)}
{\cal P}(s) \underset{s\to 0}{\sim} s^{\nu}
\qquad\mbox{with}\qquad \nu=1,2,4
\ee
in terms of the rescaled spacing defined as $s=S/\langle S\rangle_{\rm ME}$, where $S=E_{m+1}-E_m$ \cite{W59,D62a,D62b,D62c,D62d,M67,H01}.  We note that the mean spacing and density of states are related by $\langle S\rangle_{\rm ME} \, \sigma(E)=g_{\nu}$ at the energy $E=(E_{m+1}+E_m)/2$.  The forms of the surmised spacing distributions are given in Appendix~\ref{app:RMT}. 

In addition, the spectra may be characterised by the two-point correlation function defined as \cite{H01}
\be\label{Y}
Y(e-e^\prime) \equiv 1 - \Big\langle\sum_{m\ne n} \delta(e-e_m)\, \delta(e^\prime - e_n)\Big\rangle_{\rm ME}
\qquad\mbox{with}\qquad
e=g_{\nu}^{-1} \, E\, \sigma(E) \, ,
\ee
and behaving as
\begin{align}
& Y(e-e^\prime) = 1-\alpha_{\nu} \, \vert e-e^\prime\vert^{\nu} + O\left(\vert e-e^\prime\vert^{\nu+2}\right) &\mbox{for} \quad \vert e-e^\prime\vert \ll 1 \, , \label{Y-small}\\
& Y(e-e^\prime) \to 0 &\mbox{for} \quad \vert e-e^\prime\vert \gg 1 \, , \label{Y-large}
\end{align}
where
\be
\alpha_1 = \frac{\pi^2}{6} \, , \qquad
\alpha_2 = \frac{\pi^2}{3} \, , \qquad
\alpha_4 = \frac{16\, \pi^4}{135} \, .
\ee
This allows us to write
\be\label{sigma-sigma-Y}
\langle\invbreve\sigma(E)\, \invbreve\sigma(E')\rangle_{\rm ME} = g_{\nu} \sigma(E) \, \delta(E-E') + \sigma(E)\, \sigma(E') \left[ 1 - Y(e-e')\right]
\ee
with $e-e'=g_{\nu}^{-1} [E\sigma(E)-E'\sigma(E')]$.
The orthogonal, unitary, and symplectic ensembles belong to the three Wigner-Dyson universality classes by their statistical properties on small energy scales.

However, the Gaussian ensembles are not universal on large energy scales, because the average density of states is a property that is specific to each quantum system.  The Gaussian ensembles are only particular examples among a variety of other possibilities for random-matrix ensembles. Indeed, an obvious hindrance in modelling a given system in terms of random-matrix ensembles stems from the form of the average density of states, which, in general, may not be the semicircular form~(\ref{semicircle_law}) for large $N$.  In this respect, non-Gaussian ensembles have also been introduced with
\be\label{nonGaussian}
{\cal P}(\hat H) = C_{\Phi} \, {\rm e}^{- {\rm tr}\, \Phi(\hat H)} \, , 
\ee
where $\Phi(x)$ is some function growing fast enough as $\vert x\vert\to \infty$ for the probability distribution to be normalisable \cite{BZ93,BZ94}.  In principle, this function can be taylored in order for the corresponding average density of states $\sigma(E)=\langle{\rm tr}\, \delta(E-\hat H)\rangle_{\rm ME}$ to fit some experimental data.  Nevertheless, on small energy scales, the energy spectra of such general random matrices share the same universal properties of the Wigner-Dyson universality classes as the corresponding Gaussian ensembles.

%%%%%%%%%%%%%%%%%%%%%%%%%%%%%%%%%%%%%%%%%%%%%%%%%
\section{Response theory}
\label{sec:response}

\subsection{The general framework}

Now, the perturbation parameter is assumed to be time dependent, $\lambda=\lambda(t)$, and the system is initially prepared in the canonical thermal equilibrium macrostate with the inverse temperature $\beta=(k_{\rm B}T)^{-1}$.  We consider the response of the time-dependent perturbation on the mean value of some observable $\hat A^{(P)}$ at time $t$.  This mean value can be expressed as
\be
\langle \hat A^{(P)} \rangle_t= {\rm tr} \, \hat\rho_t^{(P)} \, \hat A^{(P)} \, ,
\qquad\mbox{where}\qquad
\hat\rho_t^{(P)} = \hat U_t^{(P)} \, \hat\rho_0^{(P)} \, \hat U_t^{(P) \dagger}
\ee
is the statistical operator at time $t$ expressed in terms of the unitary evolution operator defined by the following time-ordered exponential,
\be
\hat U_t^{(P)} = {\mathbb T} \exp\left[-\frac{\rm i}{\hbar} \int_0^t d\tau \, \hat H^{(P)}(\tau)\right] ,
\ee
and the initial statistical operator
\be\label{rho0-P}
\hat\rho_0^{(P)} = \frac{{\rm e}^{-\beta \hat H_0^{(P)}}}{{\rm tr}\, {\rm e}^{-\beta \hat H_0^{(P)}}} = \prod_{p=1}^P \hat\rho_{0p}
\qquad\mbox{with}\qquad
\hat\rho_{0p} = \frac{{\rm e}^{-\beta\hat H_{0p}}}{{\rm tr}\, {\rm e}^{-\beta\hat H_{0p}}} \, .
\ee
In the high-temperature limit ($\beta=0$), the statistical operator of each part becomes equal to $\hat\rho_{0p}=\hat I/N$ and, in the low-temperature limit ($\beta=\infty$), to $\hat\rho_{0p}=\vert E_{{\rm min},p}\rangle\langle E_{{\rm min},p}\vert$, where $\vert E_{{\rm min},p}\rangle$ denotes the lowest-energy eigenstate of the matrix $\hat H_{0p}$.
\\

Using time-dependent perturbation theory, we can obtain the following result,
\bea
\langle \hat A^{(P)} \rangle_t &=& \langle \hat A^{(P)} \rangle_0 + \int_0^t dt_1 \, \lambda(t_1) \, R_{AV}^{(P)}(t-t_1) 
+ \int_0^t dt_1 \int_0^{t_1} dt_2 \, \lambda(t_1) \, \lambda(t_2) \, R_{AVV}^{(P)}(t-t_1,t-t_2) \nonumber\\
&& + \int_0^t dt_1 \int_0^{t_1} dt_2 \int_0^{t_2} dt_3 \, \lambda(t_1) \, \lambda(t_2) \, \lambda(t_3) \, R_{AVVV}^{(P)}(t-t_1,t-t_2,t-t_3) + O(\lambda^4) \, ,
\label{response-A}
\eea
where the response functions of the total system are given by
\bea
R_{AV}^{(P)}(t-t_1) &\equiv& \ \; \frac{1}{{\rm i}\hbar} \ \ \, {\rm tr} \, \hat\rho_0^{(P)} \Big[ \hat A^{(P)}(t), \hat V^{(P)}(t_1) \Big] \, , \label{R1P}\\
R_{AVV}^{(P)}(t-t_1,t-t_2) &\equiv& \frac{1}{({\rm i}\hbar)^2} \; {\rm tr} \, \hat\rho_0^{(P)} \Big[\Big[ \hat A^{(P)}(t), \hat V^{(P)}(t_1) \Big], \hat V^{(P)}(t_2) \Big]\, , \label{R2P}\\
R_{AVVV}^{(P)}(t-t_1,t-t_2,t-t_3) &\equiv& \frac{1}{({\rm i}\hbar)^3} \; {\rm tr} \, \hat\rho_0^{(P)} \Big[\Big[\Big[ \hat A^{(P)}(t), \hat V^{(P)}(t_1) \Big], \hat V^{(P)}(t_2) \Big], \hat V^{(P)}(t_3) \Big] \, , \label{R3P}\\
& \vdots & \nonumber
\eea
as expressed in terms of the time-dependent operators
\be
\hat X^{(P)}(t) \equiv {\rm e}^{\frac{\rm i}{\hbar}\hat H_0^{(P)}t} \, \hat X^{(P)} \, {\rm e}^{-\frac{\rm i}{\hbar}\hat H_0^{(P)}t} \, ,
\ee
which are defined after evolution over time $t$ under the unperturbed Hamiltonian $\hat H_0^{(P)}$.  The response functions only depend on the differences of times $t-t_j$, because the time evolution operator of the unperturbed Hamiltonian commutes with the canonical equilibrium statistical operator~(\ref{rho0-P}).  We note that the response functions are given by Eqs.~(\ref{R1P}), (\ref{R2P}), (\ref{R3P}),... for $0<t_n<\cdots < t_3 < t_2 < t_1 < t$.  Otherwise, they are equal to zero in order to satisfy causality, which could be expressed by multiplying them with appropriate Heaviside functions defined as $\theta(t_j-t_{j+1})=1$ if $t_j>t_{j+1}$ and zero otherwise.

In the following, we shall focus on the response properties of the perturbation itself, i.e.,
\be\label{A=V}
\hat A^{(P)} = \hat V^{(P)}  \, .
\ee

\subsection{Statistical averages over random-matrix ensembles}

Since the different parts of the total system do not have mutual interaction because of the forms~(\ref{dfn-H0-P}) and~(\ref{dfn-V-P}) for the unperturbed Hamiltonian and the perturbation, the response functions of the total system for the observable~(\ref{A=V}) can be reduced to the sum of the response functions of each part of the system.  For instance, the first-order response function~(\ref{R1P}) for $\hat A^{(P)} = \hat V^{(P)}$ can be reduced as follows,
\bea
R_{VV}^{(P)}(t-t_1) &=& \frac{1}{{\rm i}\hbar}\, {\rm tr} \, \hat\rho_0^{(P)} \Big[ \hat V^{(P)}(t), \hat V^{(P)}(t_1) \Big] \\
&=& \frac{1}{{\rm i}\hbar}\, {\rm tr} \, \prod_{p=1}^P \hat\rho_{0p} \Big[ \sum_{p'=1}^P \hat V_{p'}(t), \sum_{p''=1}^P\hat V_{p''}(t_1) \Big] \\
&=& \frac{1}{{\rm i}\hbar}\,  \sum_{p=1}^P {\rm tr} \, \hat\rho_{0p} \Big[ \hat V_{p}(t), \hat V_{p}(t_1) \Big] \, ,
\eea
because the commutators are equal to zero for $p'\ne p''$.  In the limit of a large number of parts, the sum is carried out over randomly distributed values, which can be expressed by a statistical average using the probability distributions of the matrix ensemble according to
\be
R_{VV}^{(P)}(t-t_1) \underset{P\to\infty}{\simeq} \frac{P}{{\rm i}\hbar}\, \Big\langle {\rm tr} \, \hat\rho_{0} \Big[ \hat V(t), \hat V(t_1) \Big]  \Big\rangle_{\rm ME}\, .
\ee
In the large-system limit, the response functions can thus be obtained in terms of averages over the matrix ensemble defining these random quantum systems.  Similar results hold for higher-order response functions.

Therefore, we may introduce the mean response functions defined by the average over the matrix ensemble.  The mean first-order response function is defined by
\be\label{MRF1}
R_{VV}(t-t_1) \equiv \lim_{P\to\infty} \frac{1}{P} \, R_{VV}^{(P)}(t-t_1) = \frac{1}{{\rm i}\hbar}\, \Big\langle {\rm tr} \, \hat\rho_{0} \Big[ \hat V(t), \hat V(t_1) \Big]  \Big\rangle_{\rm ME}\, ,
\ee
while the mean $n^{\rm th}$-order response function is given by
\bea
R_{\underbrace{\scriptstyle VV\cdots V}_{n+1}}(t-t_1,t-t_2,\dots,t-t_n) &\equiv& \lim_{P\to\infty} \frac{1}{P} \, R_{\underbrace{\scriptstyle VV\cdots V}_{n+1}}^{(P)}(t-t_1,t-t_2,\dots,t-t_n) \nonumber\\
&=& \frac{1}{({\rm i}\hbar)^n}\, \Big\langle {\rm tr} \, \hat\rho_{0} \Big[\Big[\cdots \Big[\Big[ \hat V(t), \hat V(t_1) \Big],\hat V(t_2) \Big],\cdots\Big], \hat V(t_n)\Big] \Big\rangle_{\rm ME}\, .
\label{MRFn}
\eea
We note that the mean response functions of even order $n$ are equal to zero for the Gaussian ensembles~(\ref{P(V)}), because the odd moments of $V_{kl}$ are zero for such centered Gaussian distributions.  In particular, we have that $R_{VVV}(t-t_1,t-t_2) =0$.  Hence, the next-to-leading mean response function is the third-order one with $n=3$ in Eq.~(\ref{MRFn}) after the leading first-order mean response function~(\ref{MRF1}).

In the limit $P\to\infty$, the quantum system is infinite and the matrix-ensemble averages simulate a continuous spectrum of frequencies.  Accordingly, the response functions decay to zero for long-time separations between the different times, i.e., for $\vert t_{j}-t_{j+1}\vert\to\infty$ with $j=0,1,2,\dots,n-1$ and $t_0=t$.  The infinite quantum system thus behaves as mixing dynamical systems \cite{T83}.  The way the response functions decay to zero is an important issue that will be investigated in Sec.~\ref{sec:1st-resp} for the first-order mean response function. In relation to this issue, we may wonder what is the relative magnitude of universal to nonuniversal features as $N\to\infty$, which will be studied in the case of the first-order mean response function in Sec.~\ref{sec:1st-resp} and further discussed in Sec.~\ref{sec:conclusion}.

In Sec.~\ref{sec:higher-resp}, higher-order mean response functions are obtained, in particular, using the diagrammatic method.  In Sec.~\ref{sec:impulsive}, these results are applied to understand the dynamical response to impulsive perturbations.

The dependence on the number $N$ of energy levels in each part of the system is another important issue.  For $N=2$, the parts are two-level systems, which each has an oscillatory dynamics at the frequency of the spacing between the two levels, and averaging over the ensemble generates the decay to zero of the mean response functions.  In the limit $N\to\infty$, the response functions already manifest an infinite-system behaviour over the time scale $t\lesssim \hbar/\langle S\rangle_{\rm ME}=\hbar\sigma(E)$, because the dynamics of each part becomes self-averaging in this limit.  However, the response functions of the individual members of the ensemble do not decay to zero, but they present fluctuations for $t\gtrsim \hbar/\langle S\rangle_{\rm ME}=\hbar\sigma(E)$.  These fluctuations will be investigated in Sec.~\ref{sec:fluctuations}.

%%%%%%%%%%%%%%%%%%%%%%%%%%%%%%%%%%%%%%%%%%%%%%%%%
\section{First-order response and time correlation functions}
\label{sec:1st-resp}

\subsection{The two-time correlation function}

\subsubsection{Definition and generalities}

Let us define the two-time correlation function as
\be\label{dfn-correl}
C(t) = {\rm tr}\, \hat\rho_0 \, \hat V(0) \, \hat V(t) = {\rm tr}\, \hat\rho_0 \, \hat V(\tau) \, \hat V(\tau+t) \, ,
\ee
where
\be\label{rho0}
\hat\rho_0  = \frac{{\rm e}^{-\beta\hat H_0}}{Z(\beta)}
\ee
is the canonical statistical operator of the equilibrium macrostate at the inverse temperature $\beta=(k_{\rm B}T)^{-1}$,
\be\label{Z}
Z(\beta) = {\rm tr}\, {\rm e}^{-\beta \hat H_0}
\ee
is the corresponding partition function, and
\be\label{V(t)}
\hat V(t) = {\rm e}^{\frac{\rm i}{\hbar} \hat H_0 t} \, \hat V \, {\rm e}^{-\frac{\rm i}{\hbar} \hat H_0 t}
\ee
is the time-evolved perturbation operator.
Since the statistical operator commutes with the unperturbed Hamiltonian $\hat H_0$, the correlation function~(\ref{dfn-correl}) only depends on the difference between the two times.  It has the property that $C^*(t)=C(-t)$ and it satisfies the Kubo-Martin-Schwinger relation $C(t)=C({\rm i}\beta\hbar - t)$ \cite{K57,MS59}. Moreover, it can be decomposed onto the basis of eigenstates of the Hamiltonian $\hat H_0$ according to
\be\label{correl-basis}
C(t) = \sum_{k,l}  \frac{{\rm e}^{-\beta E_k}}{Z(\beta)} \, \vert V_{kl}\vert^2 \, {\rm e}^{\frac{\rm i}{\hbar} (E_l-E_k)t} \, ,
\ee
so that its real and imaginary parts are respectively given by
\be\label{Re-Im-correl}
\left\{
\begin{array}{l}
\displaystyle C_{\rm r}(t) \equiv {\rm Re}\, C(t) = \frac{1}{2\, Z(\beta)} \sum_{k,l} \left({\rm e}^{-\beta E_k}+{\rm e}^{-\beta E_l}\right) \, \vert V_{kl}\vert^2 \,\cos\left( \frac{E_l-E_k}{\hbar} \, t \right) , \\
\displaystyle C_{\rm i}(t) \equiv {\rm Im}\, C(t) = \frac{1}{2\, Z(\beta)} \sum_{k,l} \left({\rm e}^{-\beta E_k}-{\rm e}^{-\beta E_l}\right) \,  \vert V_{kl}\vert^2 \,\sin\left( \frac{E_l-E_k}{\hbar} \, t \right) , \\
\end{array}
\right.
\ee
which shows in particular that the real and imaginary parts are respectively even and odd under time reversal.  

In Eq.~(\ref{dfn-correl}), the unperturbed Hamiltonian $\hat H_0$ and the perturbation operator $\hat V$ are random matrices belonging to the statistical ensembles of the probability distributions~(\ref{P(H0)}) and~(\ref{P(V)}), so that the correlation function itself is randomly distributed.

Now, the mean first-order response function~(\ref{MRF1}) can be expressed as follows in terms of the imaginary part of the correlation function~(\ref{dfn-correl}) averaged over the matrix ensemble,
\be\label{MRF1-correl}
R_{VV}(t-t_1) = -\frac{2}{\hbar}\, \langle\,{\rm Im}\, C(t-t_1)\,\rangle_{\rm ME} \, .
\ee
Using Eq.~(\ref{var-V}), the average of the correlation function can be reduced to
\be
\langle C(t)\rangle_{\rm ME} = \frac{\gamma_\nu}{2 a_V} + \frac{1}{2 a_V} \left\langle\frac{{\rm tr}\, {\rm e}^{-\left(\beta + \frac{\rm i}{\hbar} t\right)\hat H_0} \; {\rm tr}\, {\rm e}^{\frac{\rm i}{\hbar} \hat H_0 t}}{{\rm tr}\, {\rm e}^{-\beta\hat H_0}}\right\rangle_{\rm ME} \, ,
\ee
which can equivalently be expressed as
\be\label{Mcorrel-Z}
\langle C(t)\rangle_{\rm ME} = \frac{\gamma_\nu}{2 a_V} + \frac{1}{2 a_V} \left\langle\frac{1}{Z(\beta)}\, Z\Big(\beta + \frac{\rm i}{\hbar} t\Big) \, Z\Big(-\frac{\rm i}{\hbar}t\Big)\right\rangle_{\rm ME}
\ee
in terms of the partition function~(\ref{Z}).  The latter can be written as $Z(x) = \int dE \, \invbreve\sigma(E) \, {\rm e}^{-x E}$, where $\invbreve\sigma(E)={\rm tr}\, \delta(E-\hat H_0)=\sum_k \delta(E-E_k)$ is the density of states of the unperturbed Hamiltonian.

In the following, we shall determine the time dependence of the mean correlation function on different time scales and temperature regimes.

%%%%%%%%%%%%%%%%%%%%%%%%%%%%%%%%%%%%%%%%%%%%%%%%%%%%%%%%%%
\begin{figure}[h!]\centering
{\includegraphics[width=0.5\textwidth]{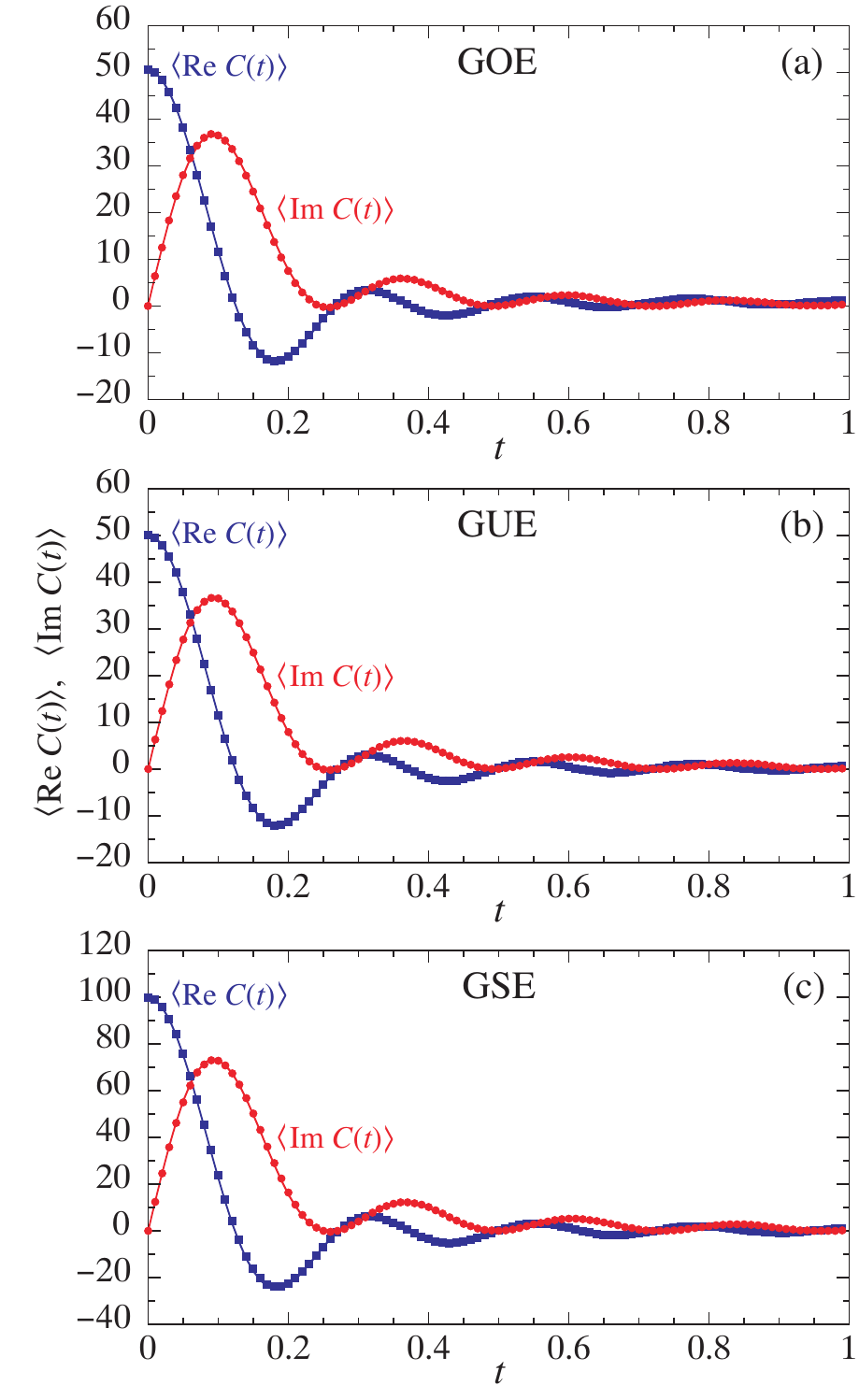}}
\caption[] {Mean correlation function~(\ref{Mcorrel-Z}) versus time for (a) GOE, (b) GUE, and (c) GSE $100\times 100$ matrices with $a_{H_0}=a_V=1$ at inverse temperature $\beta=1$. The real part is depicted by squares and the imaginary part by circles.  The statistics is carried out over $10^3$ realisations.  The time steps are $\Delta t = 0.01$. For comparison, the solid lines show the theoretical predictions by Eq.~(\ref{Mcorrel-Z-intermediate}) in terms of Bessel functions.}
\label{Fig1}
\end{figure}
%%%%%%%%%%%%%%%%%%%%%%%%%%%%%%%%%%%%%%%%%%%%%%%%%%%%%%%%%%
\begin{figure}[h!]\centering
{\includegraphics[width=0.5\textwidth]{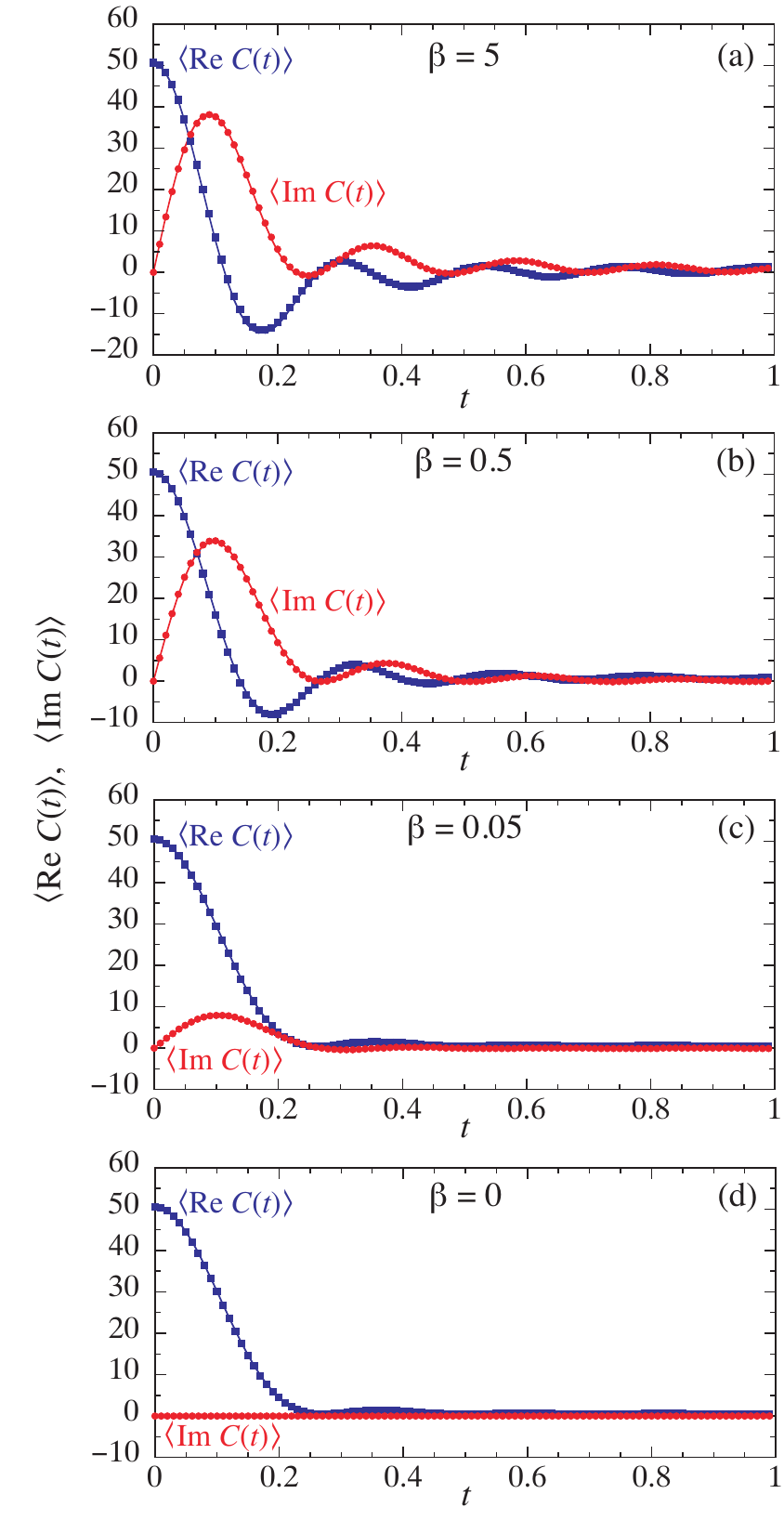}}
\caption[] {Mean correlation function~(\ref{Mcorrel-Z}) versus time for GOE $100\times 100$ matrices with $a_{H_0}=a_V=1$ at inverse temperatures (a) $\beta=5$, (b) $\beta=0.5$, (c) $\beta=0.05$, and (d) $\beta=0$. The real part is depicted by squares and the imaginary part by circles.  The statistics is carried out over $10^3$ realisations. The time steps are $\Delta t = 0.01$. For comparison, the solid lines show the theoretical predictions by Eq.~(\ref{Mcorrel-Z-intermediate}) in terms of Bessel functions.}
\label{Fig2}
\end{figure}
%%%%%%%%%%%%%%%%%%%%%%%%%%%%%%%%%%%%%%%%%%%%%%%%%%%%%%%%%%

\subsubsection{Early-time behaviour}

Since $Z(0)=g_\nu N$, the mean correlation function~(\ref{Mcorrel-Z}) at time zero ($t=0$) is given by
\be\label{Mcorrel0}
\langle C(0)\rangle_{\rm ME} = \frac{\gamma_\nu + g_\nu N}{2 a_V} \, .
\ee
Using Eq.~(\ref{var-V}) providing the values of $\gamma_\nu$ in the different ensembles, we find
\be\label{correl-time=0}
\langle C(0)\rangle_{\rm GOE} = \frac{N+1}{2 a_V} \, , \qquad
\langle C(0)\rangle_{\rm GUE} = \frac{N}{2 a_V} \, , \qquad
\langle C(0)\rangle_{\rm GSE} = \frac{4N-1}{4 a_V} \, .
\ee

Now, the early-time behaviour can be obtained by considering the Taylor expansion of the mean correlation function~(\ref{Mcorrel-Z}) in powers of the time $t$, giving
\be\label{Re-Im-Mcorrel-early}
\left\{
\begin{array}{ll}
\displaystyle \langle{\rm Re}\, C(t)\rangle_{\rm ME} = \frac{\gamma_\nu + g_\nu N}{2 a_V}  - \frac{g_\nu N K_2 }{4 a_V \hbar^2}\, t^2+ O(t^4) 
&\displaystyle \quad\mbox{with} \qquad
K_2 = \left\langle \frac{Z''(\beta)}{Z(\beta)} + \frac{Z''(0)}{Z(0)}  - 2 \frac{Z'(\beta)}{Z(\beta)}  \frac{Z'(0)}{Z(0)} \right\rangle_{\rm ME} , \\[4mm]
\displaystyle \langle{\rm Im}\, C(t)\rangle_{\rm ME} = \frac{g_\nu N K_1}{2 a_V \hbar}\, t + O(t^3) 
&\displaystyle \quad\mbox{with} \qquad
K_1 = \left\langle \frac{Z'(\beta)}{Z(\beta)} - \frac{Z'(0)}{Z(0)} \right\rangle_{\rm ME} , \\
\end{array}
\right.
\ee
where $Z'(\beta)=\partial_\beta Z(\beta)$ and $Z''(\beta)=\partial_\beta^2 Z(\beta)$.

\subsubsection{Intermediate-time behaviour}

If the size of the random matrices is large ($N\gg 1$), the self-averaging property gives the following approximation for the mean correlation function~(\ref{Mcorrel-Z}),
\be\label{Mcorrel-Z-approx}
\langle C(t)\rangle_{\rm ME} \simeq \frac{\gamma_\nu}{2 a_V} + \frac{1}{2 a_V{\mathscr Z}(\beta)}\, {\mathscr Z}\Big(\beta + \frac{\rm i}{\hbar} t\Big) \, {\mathscr Z}\Big(-\frac{\rm i}{\hbar}t\Big) \, ,
\ee
which can be evaluated using the average density of states~(\ref{semicircle_law}) for the unperturbed Hamiltonian of parameter $a_{H_0}$.  The average density of states extends over the energy range $-E_{\rm b}\leq E \leq +E_{\rm b}$ with $E_{\rm b}=\sqrt{\frac{2N}{a_{H_0}}}$, which reads $E_{\rm b}=\hbar\omega_{\rm b}$ with the frequency $\omega_{\rm b}=\sqrt{\frac{2N}{a_{H_0}\hbar^2}}$.  The matrix-ensemble averages of the partition functions can thus be evaluated as
\bea
{\mathscr Z}(\beta) &\equiv& \langle Z(\beta)\rangle_{\rm ME} = \int dE \, \sigma(E) \, {\rm e}^{-\beta E} = 2 g_\nu N \, \frac{I_1(\beta\hbar\omega_{\rm b})}{\beta\hbar\omega_{\rm b}} \, , \label{Z(b)}\\
{\mathscr Z}\Big(\beta + \frac{\rm i}{\hbar} t\Big) &\equiv&\left\langle Z\Big(\beta + \frac{\rm i}{\hbar} t\Big)\right\rangle_{\rm ME} = \int dE \, \sigma(E) \, {\rm e}^{-\beta E} \, {\rm e}^{-\frac{\rm i}{\hbar} E t} = 2 g_\nu N \, \frac{J_1\left[\omega_{\rm b}(t-{\rm i}\hbar\beta)\right]}{\omega_{\rm b}(t-{\rm i}\hbar\beta)} \, , \label{Z(b+it)}\\
{\mathscr Z}\Big(-\frac{\rm i}{\hbar} t\Big) &\equiv&\left\langle Z\Big(-\frac{\rm i}{\hbar} t\Big)\right\rangle_{\rm ME} = \int dE \, \sigma(E) \, {\rm e}^{\frac{\rm i}{\hbar} E t} = 2 g_\nu N \, \frac{J_1(\omega_{\rm b}t)}{\omega_{\rm b}t} \, , \label{Z(it)}
\eea
where $J_1(z)$ is the first-order Bessel function and $I_1(z)=-{\rm i} J_1({\rm i}z)$ the corresponding modified function \cite{AS72,GR80}.
Accordingly, we obtain
\be\label{Mcorrel-Z-intermediate}
\langle C(t)\rangle_{\rm ME} \simeq \frac{\gamma_\nu}{2 a_V} +\frac{g_\nu N \beta\, \hbar}{a_V\,\omega_{\rm b}\, t \, (t-{\rm i}\hbar\beta)} \, 
\frac{J_1\left[\omega_{\rm b}(t-{\rm i}\hbar\beta)\right]\, J_1(\omega_{\rm b}t)}{I_1(\beta\hbar\omega_{\rm b})} \, ,
\ee
which shows that the characteristic time scale of the response is inversely proportional to the bandwith, $\tau_{\rm b}=2\pi/\omega_{\rm b}$.
\\

Since $I_1(z)=z/2+O(z^3)$ and $J_1(z)=z/2+O(z^3)$, the limit $t\to 0$ of Eq.~(\ref{Mcorrel-Z-intermediate}) gives the same value $\langle C(0)\rangle_{\rm ME} =(\gamma_\nu + g_\nu N)/(2 a_V)$ as Eq.~(\ref{Mcorrel0}).  In this regard, the approximation, which consists in factorizing the average of a product into a product of averages in Eq.~(\ref{Mcorrel-Z-approx}), is consistent with the exact value at time $t=0$.
\\

Figure~\ref{Fig1} shows that Eq.~(\ref{Mcorrel-Z-intermediate}) gives an excellent approximation for the intermediate-time behaviour of the mean correlation function in the GOE, GUE, and GSE.  The early-time behaviour is well described by Eq.~(\ref{Re-Im-Mcorrel-early}).  Otherwise, there is a very similar dependence on time for the different ensembles.  
\\

The dependence of the mean correlation function on the inverse temperature $\beta$ is illustrated in Fig.~\ref{Fig2} in the GOE.  The oscillatory behaviour is more pronounced at lower than higher temperatures.  Although the real part of the mean correlation function always decays from the same zero-time value~(\ref{correl-time=0}), the magnitude of its imaginary part decreases as the temperature increases.  In the limit of infinitely high temperature (i.e., $\beta=0$), the imaginary part becomes equal to zero.

\subsubsection{Long-time behaviour}

To investigate the long-time behaviour of the mean correlation function, it is convenient to separate the double sum in Eq.~(\ref{correl-basis}) into the sum over the terms with $k=l$ and the sum over the other terms with $k\ne l$.  Moreover, to deal with the case of symplectic systems, we need to write the label $k$ of the eigenstates as the label $m$ of the eigenvalue $E_k=E_m$ and the index $i \in\{ 1,g_\nu \}$ of multiplicity.  We thus set $k=(m,i)$ and $l=(n,j)$.  In this way, the real and imaginary parts of the mean correlation function can be written in the following forms,
\be\label{Re-Im-Mcorrel-levels}
\left\{
\begin{array}{ll}
\displaystyle \langle{\rm Re}\, C(t)\rangle_{\rm ME} = \frac{\gamma_\nu + g_\nu}{2 a_V}  +\frac{g_\nu}{2 a_V} \left\langle \sum_{m\ne n} \frac{{\rm e}^{-\beta E_m}}{\tilde Z(\beta)} \, \cos\left( \frac{E_n-E_m}{\hbar} \, t \right)\right\rangle_{\rm ME} , \\[4mm]
\displaystyle \langle{\rm Im}\, C(t)\rangle_{\rm ME} = \frac{g_\nu}{2 a_V}\, \left\langle \sum_{m\ne n} \frac{{\rm e}^{-\beta E_m}}{\tilde Z(\beta)} \, \sin\left( \frac{E_n-E_m}{\hbar} \, t \right)\right\rangle_{\rm ME} , \\
\end{array}
\right.
\ee
where $\tilde Z(\beta)=\sum_m {\rm e}^{-\beta E_m}=Z(\beta)/g_\nu$.

To obtain the long-time behaviour, we may introduce the time average defined as
\be
\overline{F(t)} = \lim_{{\mathscr T}\to\infty} \frac{1}{\mathscr T} \int_0^{\mathscr T} F(t) \, dt
\ee
for some function $F(t)$ depending on time.
Since
\be\label{time_aver-cos-sin}
\overline{\cos\left( \frac{E_n-E_m}{\hbar} \, t \right)} = \delta_{mn}
\qquad\mbox{and}\qquad
\overline{\sin\left( \frac{E_n-E_m}{\hbar} \, t \right)} = 0 \, ,
\ee
we have that
\be\label{Re-Im-Mcorrel-infty}
\left\{
\begin{array}{ll}
\displaystyle \lim_{t\to\infty}\langle{\rm Re}\, C(t)\rangle_{\rm ME} =\langle\overline{{\rm Re}\, C(t)}\rangle_{\rm ME}= \frac{\gamma_\nu + g_\nu}{2 a_V} \, , \\[4mm]
\displaystyle \lim_{t\to\infty}\langle{\rm Im}\, C(t)\rangle_{\rm ME} = \langle\overline{{\rm Im}\, C(t)}\rangle_{\rm ME} =  0 , \\
\end{array}
\right.
\ee
so that the real part of the mean correlation function converges respectively towards the following positive constants,
\be
\langle\overline{{\rm Re}\, C(t)}\rangle_{\rm GOE}= \frac{1}{a_V} \, , \qquad
\langle\overline{{\rm Re}\, C(t)}\rangle_{\rm GUE}= \frac{1}{2 a_V} \, , \qquad
\langle\overline{{\rm Re}\, C(t)}\rangle_{\rm GSE}= \frac{3}{4 a_V} \, , 
\ee
while its imaginary part giving the first-order response function decay to zero, as aforementioned.
\\

Next, the issue is to determine if the mean correlation function has an algebraic or faster decay to its asymptotic value.  Since the long-time limit probes the small energy scale of the spectrum, we may expect that the universal features of the energy spectrum will manifest themselves in the long-time behaviour of the mean correlation function and, in particular, that the long-time decay is controlled by the spacing distribution~(\ref{P(s)}).  In order to investigate this issue, we express the mean correlation function as
\be\label{Mcorrel-sigmas}
\langle C(t)\rangle_{\rm ME} = \frac{\gamma_\nu}{2 a_V} + \frac{1}{2 a_V} \int dE \int dE' \, {\rm e}^{-\beta E} \, {\rm e}^{\frac{\rm i}{\hbar} (E'-E)t} \left\langle\frac{\invbreve\sigma(E)\, \invbreve\sigma(E')}{Z(\beta)}\right\rangle_{\rm ME}
\ee
in terms of the spectral density $\invbreve\sigma(E)=\sum_k\delta(E-E_k)$ of the individual members of the ensemble.  If $N$ is large enough, the statistical average can be approximated by factorizing out $\langle Z(\beta)\rangle_{\rm ME}^{-1}$ and by using Eq.~(\ref{sigma-sigma-Y}) to obtain
\be\label{Mcorrel-sigmas-approx}
\langle C(t)\rangle_{\rm ME} \simeq \frac{\gamma_\nu + g_\nu}{2 a_V} + \frac{1}{2 a_V \langle Z(\beta)\rangle_{\rm ME}} \int dE \int dE' \, {\rm e}^{-\beta E} \, {\rm e}^{\frac{\rm i}{\hbar} (E'-E)t} \, \sigma(E)\, \sigma(E') \left[ 1-Y(e-e')\right]
\ee
in terms of the two-point correlation function~(\ref{Y}) to account for the universal properties on small energy scales.
As shown in Appendix~\ref{app:long-time}, the long-time behaviour of the time-dependent mean correlation function is found to be
\be\label{Mcorrel-long-time}
\langle C(t)\rangle_{\rm ME} \simeq \frac{\gamma_\nu + g_\nu}{2 a_V} + \frac{A_\nu}{a_V} \, \Gamma(\nu+1) \, \cos\frac{\pi(\nu+1)}{2} \left[ \left(\frac{\hbar}{t}\right)^{\nu+1} + {\rm i} \, \beta \, \frac{\nu+1}{2} \left(\frac{\hbar}{t}\right)^{\nu+2} \right] + O\left(\frac{1}{t^{\nu+3}}\right)
\ee
with the positive coefficient~(\ref{coeff-A_nu}).  We note that the first two terms of this asymptotic expansion are compatible with each other, given that the two-time correlation function should satisfy the Kubo-Martin-Schwinger relation $C(t)=C({\rm i}\beta\hbar-t)$ and also $C(t)=C(-t)^*$.

In the GOE case ($\nu=1$), the result~(\ref{Mcorrel-long-time}) is thus given by
\be\label{Mcorrel-long-time-GOE}
\langle C(t)\rangle_{\rm GOE} \simeq \frac{1}{a_V} - \frac{A_1}{a_V} \left[ \left(\frac{\hbar}{t}\right)^{2} + {\rm i} \, \beta \left(\frac{\hbar}{t}\right)^{3}\right] + O\left(\frac{1}{t^4}\right) ,
\ee
so that the decay is algebraic, going as $t^{-2}$ for the real part and as $t^{-3}$ for the imaginary part, i.e., for the first-order response function.  Moreover, the latter vanishes as $\beta=(k_{\rm B}T)^{-1}$ in the limit of high temperature.  In contrast, in the GUE case ($\nu=2$), the two evaluated terms are equal to zero because $\cos\frac{\pi(\nu+1)}{2}=0$, so that the decay should be faster than $t^{-5+\delta}$ with some arbitrarily small $\delta >0$ according to Eq.~(\ref{Mcorrel-long-time}).  In the GSE case ($\nu=4$), the two evaluated terms are also equal to zero for the same reason and the decay should be faster than $t^{-7+\delta}$ with $\delta >0$.

\subsection{The two-time correlation function for two-level systems}

\subsubsection{General expression}

We confirm the previous results by considering two-level systems ($N=2$), for which analytical results can be obtained.  In this case, the real and imaginary parts of the correlation function given by Eq.~(\ref{Re-Im-Mcorrel-levels}) reduce to the following expressions involving only the spacing $S=E_2-E_1$ between the two levels,
\be\label{Re-Im-Mcorrel-levels-N=2}
\left\{
\begin{array}{ll}
\displaystyle \langle{\rm Re}\, C(t)\rangle_{\rm ME} = \frac{\gamma_\nu + g_\nu}{2 a_V}  +\frac{g_\nu}{2 a_V} \left\langle \cos \frac{S t}{\hbar} \right\rangle_{\rm ME} , \\[4mm]
\displaystyle \langle{\rm Im}\, C(t)\rangle_{\rm ME} = \frac{g_\nu}{2 a_V}\, \left\langle \tanh\frac{\beta S}{2} \, \sin\frac{St}{\hbar} \right\rangle_{\rm ME} = -\frac{g_\nu}{2 a_V}\, \tan\left(\frac{\beta\hbar}{2}\frac{\partial}{\partial t}\right) \left\langle \cos\frac{St}{\hbar} \right\rangle_{\rm ME} , \\
\end{array}
\right.
\ee
which can be directly evaluated using the spacing distribution ${\cal P}(S)$.  We note that the real part of the mean correlation function is independent of the temperature in the case of two-level systems.  The last equality in the second line of Eq.~(\ref{Re-Im-Mcorrel-levels-N=2}) is obtained by expanding the hyperbolic tangent in Taylor series, by using the identity $\langle S^{2n-1}\sin(St/\hbar)\rangle=(-1)^{n}(\hbar\partial_t)^{2n-1}\langle\cos(St/\hbar)\rangle$, and by resumming the Taylor series into minus the tangent function with its argument replaced by $\frac{1}{2}\beta\hbar\partial_t$.

An interesting case is the zero-temperature limit, where $\beta\to\infty$.  In practice, this corresponds to the limit where the thermal energy is much smaller than the mean spacing between the levels, $k_{\rm B}T\ll \langle S\rangle_{\rm ME}$.   In this limit, we have that $\lim_{\beta\to\infty}\vert\tanh(\beta S/2)\vert=1$.  As a consequence, the imaginary part of the mean correlation function is given by
\be\label{ImC-T=0}
\beta=\infty: \qquad
\langle{\rm Im}\, C(t)\rangle_{\rm ME} = \frac{g_\nu}{2 a_V}\, \left\langle \sin\frac{St}{\hbar} \right\rangle_{\rm ME} .
\ee
Remarkably, algebraic decays become possible for GUE and GSE in this limit, as shown here below.

\subsubsection{GOE two-level systems} 

\begin{figure}[h!]\centering
{\includegraphics[width=0.5\textwidth]{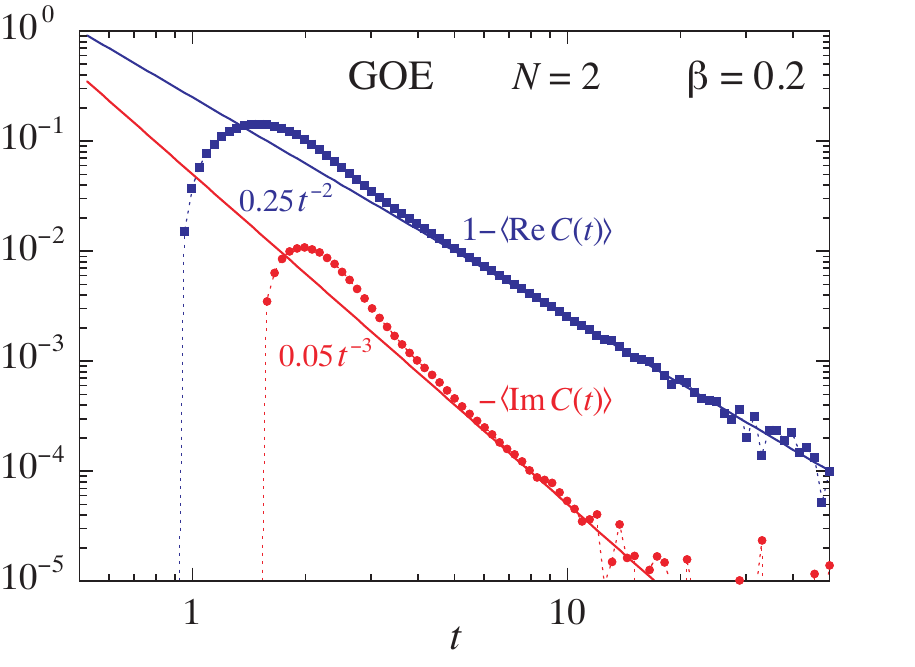}}
\caption[] {Mean correlation function~(\ref{Mcorrel-Z}) versus time for GOE $2\times 2$ matrices with $a_{H_0}=a_V=1$ at inverse temperature $\beta=0.2$. The real part is depicted by squares and the imaginary part by circles.  The statistics is carried out over $10^8$ realisations.  For comparison, the solid lines show the power-law decays predicted by Eq.~(\ref{Re-Im-Mcorrel-GOE-N=2}).}
\label{Fig3}
\end{figure}

\begin{figure}[h!]\centering
{\includegraphics[width=0.5\textwidth]{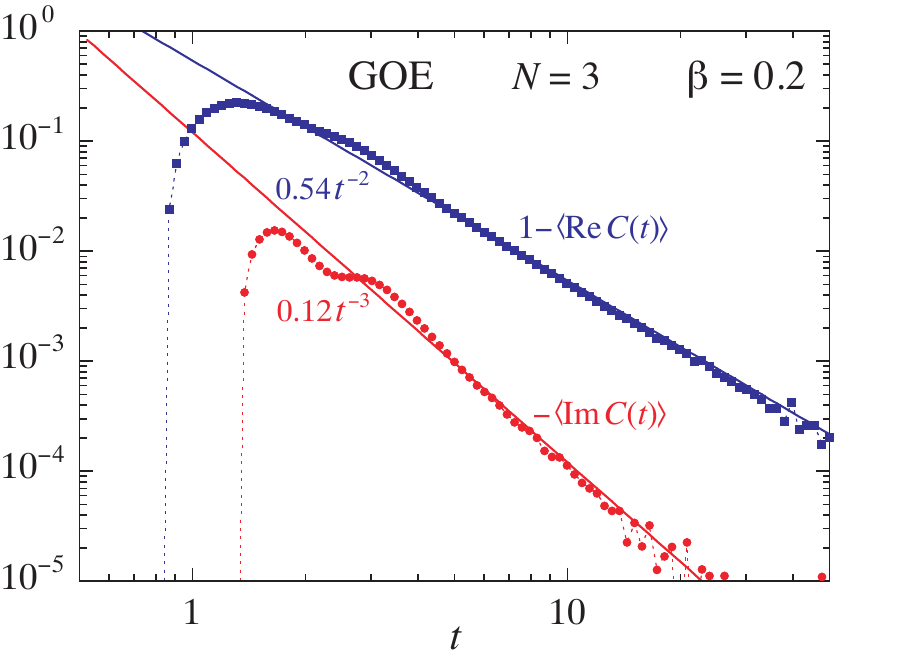}}
\caption[] {Mean correlation function~(\ref{Mcorrel-Z}) versus time for GOE $3\times 3$ matrices with $a_{H_0}=a_V=1$ at inverse temperature $\beta=0.2$. The real part is depicted by squares and the imaginary part by circles.  The statistics is carried out over $10^8$ realisations.  For comparison, the solid lines show the power-law decays predicted by Eq.~(\ref{Mcorrel-long-time-GOE}) with fitted prefactors.}
\label{Fig4}
\end{figure}

The details of the calculations to obtain the asymptotic behaviour of the mean correlation function are given in Appendix~\ref{app:GOE.N=2}, leading to
\be\label{Re-Im-Mcorrel-GOE-N=2}
\left\{
\begin{array}{ll}
\displaystyle \langle{\rm Re}\, C(t)\rangle_{\rm GOE} = \frac{1}{a_V}  -\frac{a_{H_0}\hbar^2}{4a_V\, t^2} +O\left(\frac{1}{t^4}\right) , \\[4mm]
\displaystyle \langle{\rm Im}\, C(t)\rangle_{\rm GOE} = -  \frac{\beta a_{H_0}\hbar^3}{4a_V\, t^3} +O\left(\frac{\beta}{t^5}\right) . \\
\end{array}
\right.
\ee
Accordingly, the decay of the mean correlation function towards its asymptotic value is algebraic in the GOE case.  This long-time behaviour is confirmed by numerical simulations, as shown in Fig.~\ref{Fig3}.  We note that the imaginary part is proportional to the inverse temperature $\beta$, so that the amplitude of the first-order response function tends to decrease at high temperature.
Similar algebraic decays are numerically observed for the GOE with $N=3$, as seen in Fig.~\ref{Fig4}.

In the zero-temperature limit, the imaginary part of the mean correlation function is exactly given by
\be\label{ImC-T=0.GOE}
\beta=\infty: \qquad
\langle{\rm Im}\, C(t)\rangle_{\rm GOE} = \frac{1}{2 a_V}\, \left\langle \sin\frac{St}{\hbar} \right\rangle_{\rm GOE} = \frac{1}{2 a_V}\sqrt{\frac{\pi}{a_{H_0}}} \, \frac{t}{\hbar} \, \exp\left(-\frac{t^2}{a_{H_0}\hbar^2}\right) ,
\ee
as shown in Appendix~\ref{app:GOE.N=2}, so that the asymptotic decay is here faster than algebraic in contrast to what happens for positive temperatures.

%%%%%%%%%%%%%%%%%%%%%%%%%%%%%%%%%%%%%%%%%%%%%%%%%%%%%%%%%%
%%%%%%%%%%%%%%%%%%%%%%%%%%%%%%%%%%%%%%%%%%%%%%%%%%%%%%%%%%
%%%%%%%%%%%%%%%%%%%%%%%%%%%%%%%%%%%%%%%%%%%%%%%%%%%%%%%%%%

\subsubsection{GUE two-level systems} 

\begin{figure}[h!]\centering
{\includegraphics[width=0.5\textwidth]{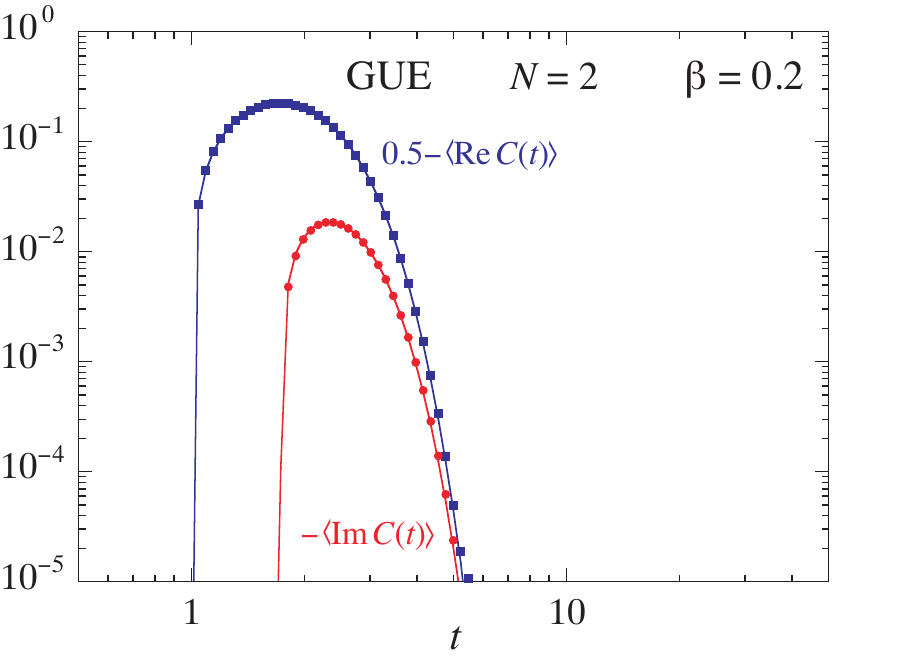}}
\caption[] {Mean correlation function~(\ref{Mcorrel-Z}) versus time for GUE $2\times 2$ matrices with $a_{H_0}=a_V=1$ at inverse temperature $\beta=0.2$. The real part is depicted by squares and the imaginary part by circles.  The statistics is carried out over $10^8$ realisations.  For comparison, the solid lines show the predictions of Eq.~(\ref{Re-Im-Mcorrel-GUE-N=2}).}
\label{Fig5}
\end{figure}

Here, the real part of the mean correlation function can be exactly calculated as shown in Appendix~\ref{app:GUE.N=2} and its imaginary part can be deduced using the second line of Eq.~(\ref{Re-Im-Mcorrel-levels-N=2}) to give
\be\label{Re-Im-Mcorrel-GUE-N=2}
\left\{
\begin{array}{ll}
\displaystyle \langle{\rm Re}\, C(t)\rangle_{\rm GUE} = \frac{1}{2a_V}  +\frac{1}{2a_V} \left(1-\frac{t^2}{a_{H_0}\hbar^2}\right)\exp\left(-\frac{t^2}{2a_{H_0}\hbar^2}\right)  , \\[4mm]
\displaystyle \langle{\rm Im}\, C(t)\rangle_{\rm GUE} = \frac{\beta}{4a_V} \frac{t}{a_{H_0}\hbar} \left(3-\frac{t^2}{a_{H_0}\hbar^2}\right)\exp\left(-\frac{t^2}{2a_{H_0}\hbar^2}\right) \left[1+O(\beta^2/a_{H_0})\right] . \\
\end{array}
\right.
\ee
Therefore, the decay is faster than any inverse power of time, which is confirmed by numerical simulations, as seen in Fig.~\ref{Fig5}.  As in the GOE case, the imaginary part is proportional to the inverse temperature $\beta$ as the direct consequence of the second line of Eq.~(\ref{Re-Im-Mcorrel-levels-N=2}).

%%%%%%%%%%%%%%%%%%%%%%%%%%%%%%%%%%%%%%%%%%%%%%%%%%%%%%%%%%
%%%%%%%%%%%%%%%%%%%%%%%%%%%%%%%%%%%%%%%%%%%%%%%%%%%%%%%%%%
\begin{figure}[h!]\centering
{\includegraphics[width=0.5\textwidth]{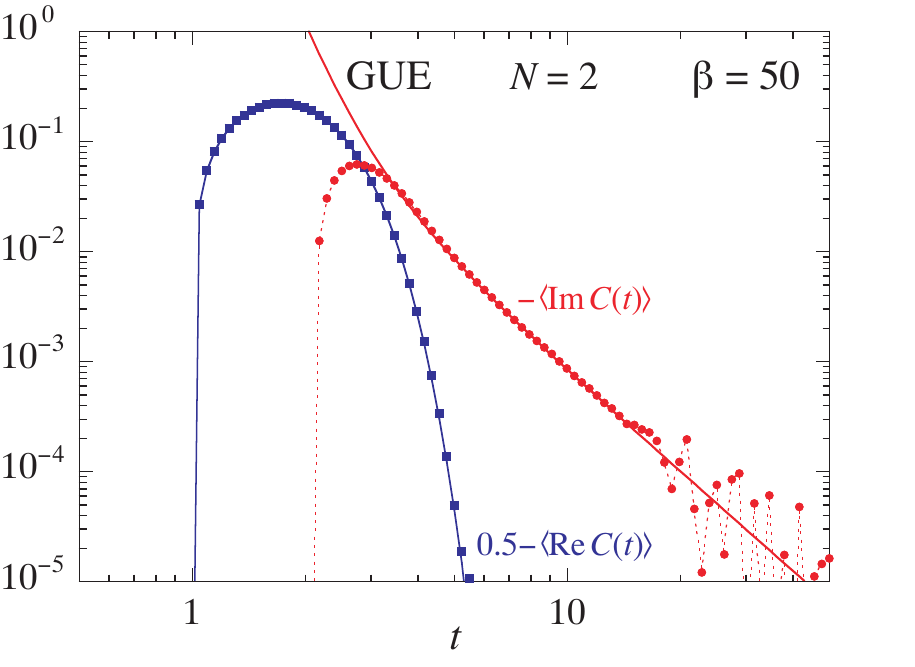}}
\caption[] {Mean correlation function~(\ref{Mcorrel-Z}) versus time for GUE $2\times 2$ matrices with $a_{H_0}=a_V=1$ at inverse temperature $\beta=50$. The real part is depicted by squares and the imaginary part by circles.  The statistics is carried out over $10^8$ realisations.  For comparison, the solid lines show the predictions of Eq.~(\ref{Re-Im-Mcorrel-GUE-N=2}) for the real part and of Eq.~(\ref{ImC-T=0.GUE}) for the imaginary part including the $t^{-9}$ term given in Eq.~(\ref{GUE-<sin>}).}
\label{Fig6}
\end{figure}
%%%%%%%%%%%%%%%%%%%%%%%%%%%%%%%%%%%%%%%%%%%%%%%%%%%%%%%%%%

In the zero-temperature limit, the imaginary part of the mean correlation function now has an algebraic decay according to
\bea\label{ImC-T=0.GUE}
\beta=\infty: \qquad
\langle{\rm Im}\, C(t)\rangle_{\rm GUE} &=& \frac{1}{2 a_V}\, \left\langle \sin\frac{St}{\hbar} \right\rangle_{\rm GUE} \nonumber\\
&=& -\frac{1}{a_V}\sqrt{\frac{2}{\pi}} \left[ \left(\frac{\hbar\sqrt{a_{H_0}}}{t}\right)^3 + 6 \left(\frac{\hbar\sqrt{a_{H_0}}}{t}\right)^5 + 45 \left(\frac{\hbar\sqrt{a_{H_0}}}{t}\right)^7 + \cdots + O\left({\rm e}^{-\frac{t^2}{2 a_{H_0}\hbar^2}}\right)\right] , \qquad
\eea
which is obtained in Appendix~\ref{app:GUE.N=2}.  Numerical computations confirm this prediction of power-law decay as shown in Fig.~\ref{Fig6}.  We here note the contrast with respect to the behaviour at positive temperatures where the decay is much faster than algebraic in Fig.~\ref{Fig5}.

\subsubsection{GSE two-level systems} 

\begin{figure}[h!]\centering
{\includegraphics[width=0.5\textwidth]{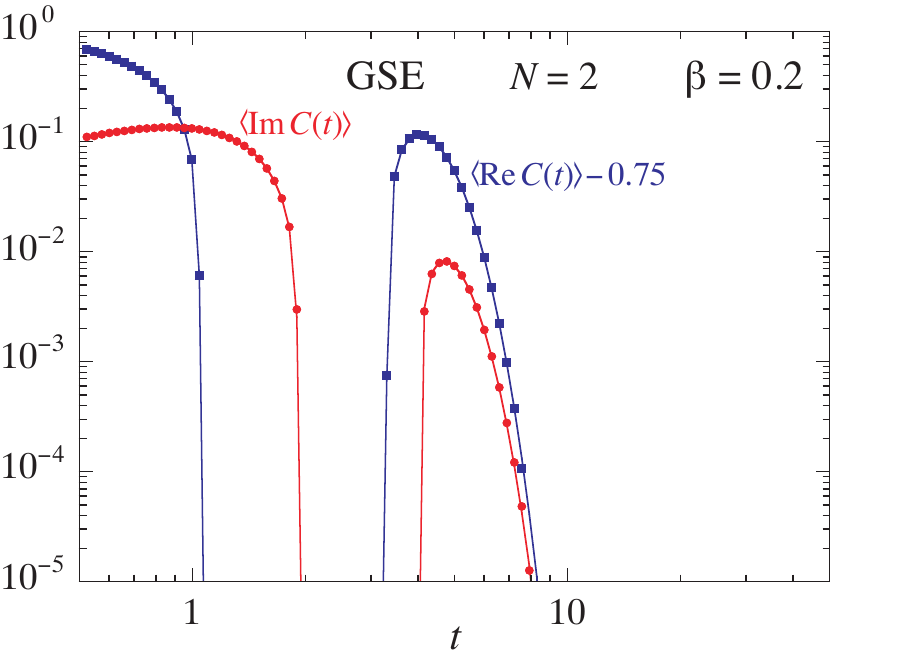}}
\caption[] {Mean correlation function~(\ref{Mcorrel-Z}) versus time for GSE $2\times 2$ matrices with $a_{H_0}=a_V=1$ at inverse temperature $\beta=0.2$. The real part is depicted by squares and the imaginary part by circles.  The statistics is carried out over $10^8$ realisations.  For comparison, the solid lines show the predictions of Eq.~(\ref{Re-Im-Mcorrel-GSE-N=2}).}
\label{Fig7}
\end{figure}

In Appendix~\ref{app:GSE.N=2}, the exact calculation is performed for the real part of the mean correlation function and the second line of Eq.~(\ref{Re-Im-Mcorrel-levels-N=2}) gives its imaginary part, so that we here have that
\be\label{Re-Im-Mcorrel-GSE-N=2}
\left\{
\begin{array}{ll}
\displaystyle \langle{\rm Re}\, C(t)\rangle_{\rm GSE} = \frac{3}{4a_V}  +\frac{1}{a_V} \left(1-\frac{t^2}{a_{H_0}\hbar^2}+\frac{t^4}{12 a_{H_0}^2\hbar^4}\right)\exp\left(-\frac{t^2}{4a_{H_0}\hbar^2}\right)  , \\[4mm]
\displaystyle \langle{\rm Im}\, C(t)\rangle_{\rm GSE} = \frac{\beta}{4a_V} \frac{t}{a_{H_0}\hbar} \left(5-\frac{5\, t^2}{3 a_{H_0}\hbar^2}+\frac{t^4}{12 a_{H_0}^2\hbar^4}\right)\exp\left(-\frac{t^2}{4a_{H_0}\hbar^2}\right) \left[1+O(\beta^2/a_{H_0})\right] . \\
\end{array}
\right.
\ee
The decay in time is thus here also faster than algebraic, as in the GUE case.  This behaviour is also confirmed by numerical simulations, as seen in Fig.~\ref{Fig7}.  As in the previous cases, the imaginary part is proportional to the inverse temperature $\beta$.

%%%%%%%%%%%%%%%%%%%%%%%%%%%%%%%%%%%%%%%%%%%%%%%%%%%%%%%%%%
%%%%%%%%%%%%%%%%%%%%%%%%%%%%%%%%%%%%%%%%%%%%%%%%%%%%%%%%%%
\begin{figure}[h!]\centering
{\includegraphics[width=0.5\textwidth]{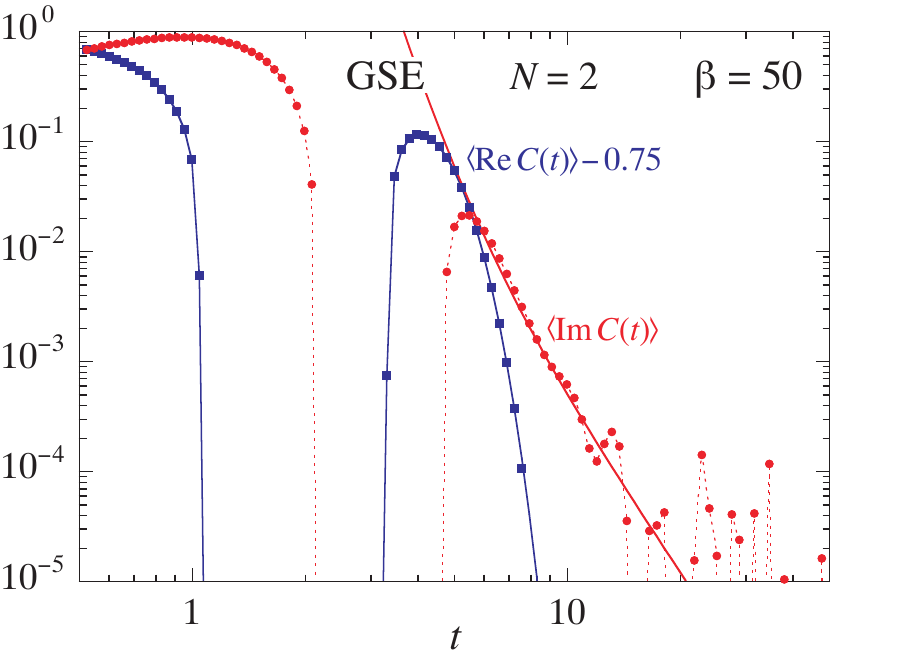}}
\caption[] {Mean correlation function~(\ref{Mcorrel-Z}) versus time for GSE $2\times 2$ matrices with $a_{H_0}=a_V=1$ at inverse temperature $\beta=50$. The real part is depicted by squares and the imaginary part by circles.  The statistics is carried out over $10^8$ realisations.  For comparison, the solid lines show the predictions of Eq.~(\ref{Re-Im-Mcorrel-GSE-N=2}) for the real part and of Eq.~(\ref{ImC-T=0.GSE}) for the imaginary part including the $t^{-11}$ term given in Eq.~(\ref{GSE-<sin>}).}
\label{Fig8}
\end{figure}
%%%%%%%%%%%%%%%%%%%%%%%%%%%%%%%%%%%%%%%%%%%%%%%%%%%%%%%%%%

In the zero-temperature limit, the imaginary part of the mean correlation function also has an algebraic decay as in the previous case, but here given by
\bea\label{ImC-T=0.GSE}
\beta=\infty: \qquad
\langle{\rm Im}\, C(t)\rangle_{\rm GSE} &=& \frac{1}{a_V}\, \left\langle \sin\frac{St}{\hbar} \right\rangle_{\rm GSE} \nonumber\\
&=& \frac{64}{a_V\sqrt{\pi}} \left[ \left(\frac{\hbar\sqrt{a_{H_0}}}{t}\right)^5 + 30 \left(\frac{\hbar\sqrt{a_{H_0}}}{t}\right)^7 + 840 \left(\frac{\hbar\sqrt{a_{H_0}}}{t}\right)^9 + \cdots + O\left({\rm e}^{-\frac{t^2}{4 a_{H_0}\hbar^2}}\right)\right] , \qquad
\eea
which is deduced in Appendix~\ref{app:GSE.N=2}.  The prediction of power-law decay is supported by the numerical computations shown in Fig.~\ref{Fig8}.  
\\

Therefore, the analysis for the two-level systems confirms that, at positive temperatures, the decay is algebraic going as $t^{-2}+{\rm i}\beta\hbar t^{-3}+O(t^{-4})$ in the GOE case, but can be significantly faster in the GUE and GSE cases.  The results obtained in Refs.~\cite{AF00a,AF00b} for the real part of the correlation functions are here recovered.  In the zero-temperature limit ($\beta=\infty$), the situation is exchanged: the imaginary part of the mean correlation functions presents an algebraic decay in the GUE and GSE cases, although its decay is faster than algebraic in the GOE case.

\subsection{The associated spectral density}

\subsubsection{Definition and generalities}

The spectral density, defined as the Fourier transform of the two-time correlation function~(\ref{dfn-correl}), is a measure of the frequency content of the system dynamics.  It is given by
\be
S(\omega) = \int_{-\infty}^{+\infty} C(t) \, {\rm e}^{{\rm i}\omega t} \, dt
= 2\pi\hbar \sum_{k,l} \frac{{\rm e}^{-\beta E_k}}{Z(\beta)} \, \vert V_{kl}\vert^2 \, \delta(\hbar\omega-E_k + E_l)
\ee
and it satisfies the following identity,
\be\label{KMS}
S(-\omega) = {\rm e}^{\beta\hbar\omega} \, S(\omega) \, ,
\ee
as a consequence of the aforementioned Kubo-Martin-Schwinger relation for the two-time correlation function.
\\

According to Eq.~(\ref{var-V}), the statistical average of the spectral density over the matrix ensemble can be expressed as
\be\label{M-spectral}
\langle S(\omega)\rangle_{\rm ME} = \frac{\pi}{a_V} \, \gamma_\nu \, \delta(\omega) + \frac{\pi\hbar}{a_V} \int dE \, {\rm e}^{-\beta E} \left\langle\frac{\invbreve\sigma(E)\, \invbreve\sigma(E-\hbar\omega)}{Z(\beta)}\right\rangle_{\rm ME}
\ee
in terms of the density of states $\invbreve\sigma(E)=\sum_k\delta(E-E_k)$.  If $N$ is large enough, we may use Eq.~(\ref{sigma-sigma-Y}) after factorizing out $\langle Z(\beta)\rangle_{\rm ME}^{-1}$, and obtain the following approximation for the mean spectral density,
\be\label{Mspctrl-Y}
\langle S(\omega)\rangle_{\rm ME} \simeq \frac{\pi}{a_V} \, (\gamma_\nu + g_\nu) \, \delta(\omega) + \frac{\pi\hbar}{a_V \langle Z(\beta)\rangle_{\rm ME}} \int dE \, {\rm e}^{-\beta E} \, \sigma(E)\, \sigma(E-\hbar\omega) \left[ 1-Y(\Delta e)\right] ,
\ee
which involves the two-point correlation function~(\ref{Y}) and
\be\label{De-omega}
\Delta e = \frac{\sigma(E)+E\, \partial_E\sigma(E)}{g_\nu} \, \hbar\omega + O(\omega^2) \, .
\ee
We note the existence of a delta peak at zero frequency in direct relation to the asymptotic constant~(\ref{Re-Im-Mcorrel-infty}) of the mean two-time correlation function.  Moreover, there is a possible manifestation of the universal properties of energy spectra because of the dependence on the two-point correlation function $Y(\Delta e)$.  
\\

\begin{figure}[h!]\centering
{\includegraphics[width=0.65\textwidth]{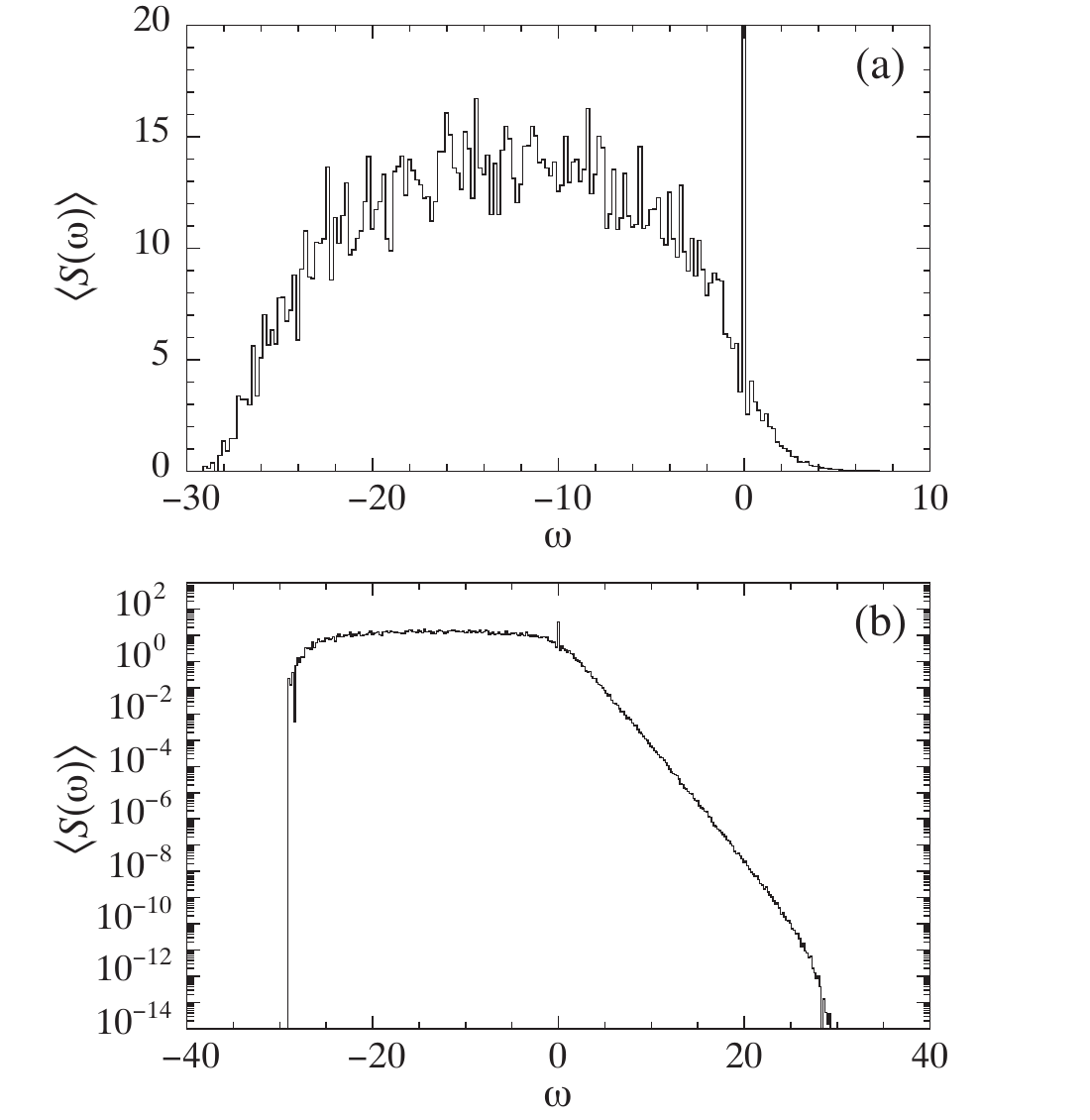}}
\caption[] {(a) Mean spectral density~(\ref{M-spectral}) versus frequency for GOE $100\times 100$ matrices with $a_{H_0}=a_V=1$ at inverse temperature $\beta=1$. The statistics is carried out over $10^2$ realisations.  The histogram is drawn with $\Delta\omega=0.2$.  (b) Same as (a) in logarithmic scale.}
\label{Fig9}
\end{figure}

\begin{figure}[h!]\centering
{\includegraphics[width=0.65\textwidth]{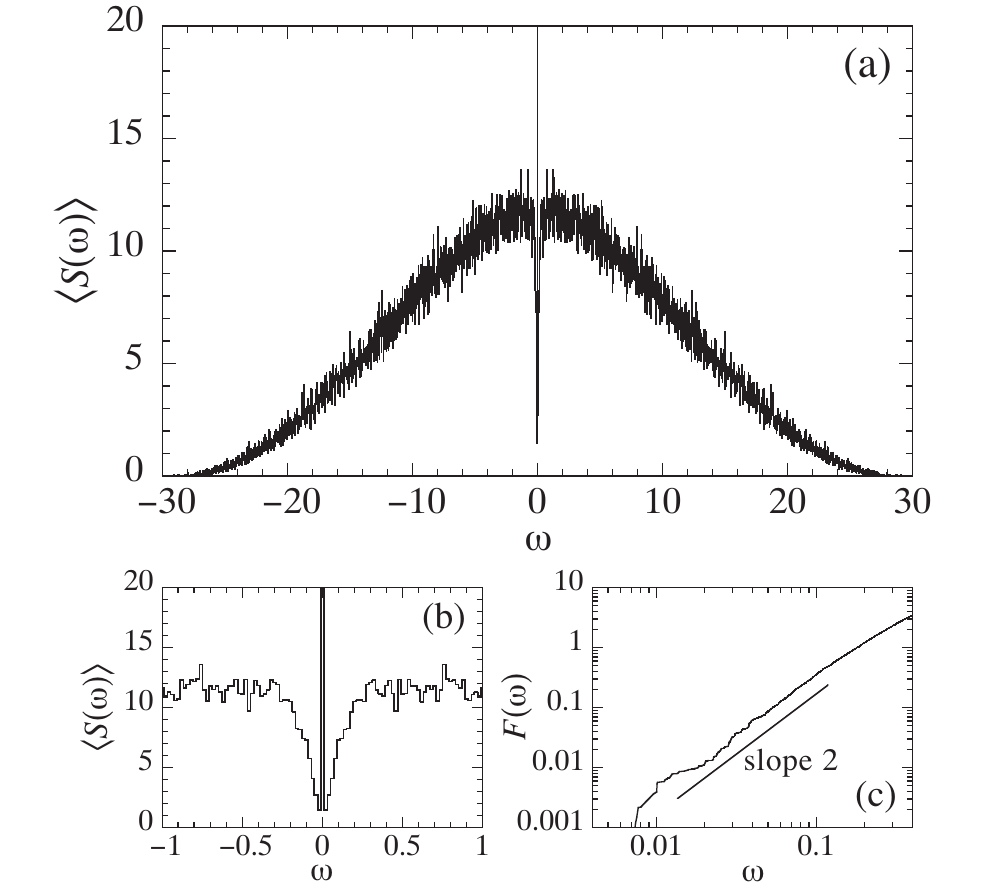}}
\caption[] {(a) Mean spectral density~(\ref{M-spectral}) versus frequency for GOE $100\times 100$ matrices with $a_{H_0}=a_V=1$ at inverse temperature $\beta=0$. The statistics is carried out over $10^2$ realisations.  The histogram is drawn with $\Delta\omega=0.02$.  (b) Zoom near zero frequency to show the dip.  (c) Log-log plot of the cumulative function~(\ref{F(w)}) to display its universal scaling as $F(\omega)\sim\omega^2$ for GOE.}
\label{Fig10}
\end{figure}

Figures~\ref{Fig9} and~\ref{Fig10} show two examples of spectral densities with their delta peak at zero temperature.  A consequence of the Kubo-Martin-Schwinger relation~(\ref{KMS}) is the apparent exponential decay seen at positive frequencies in Fig.~\ref{Fig9}(b) in relation with the lookalike plateau of the spectral density at negative frequencies.

%%%%%%%%%%%%%%%%%%%%%%%%%%%%%%%%%%%%%%%%%%%%%%%%%%%%%%%%%%
%%%%%%%%%%%%%%%%%%%%%%%%%%%%%%%%%%%%%%%%%%%%%%%%%%%%%%%%%%
%%%%%%%%%%%%%%%%%%%%%%%%%%%%%%%%%%%%%%%%%%%%%%%%%%%%%%%%%%

\subsubsection{Large-frequency behaviour}

For frequencies larger than the mean spacing, i.e., for $\vert\hbar\omega\vert \gg \langle E_{m+1}-E_m\rangle_{\rm ME}$, the delta peak does not contribute and the two-point correlation function vanishes, $Y(\Delta e)\simeq 0$, so that the mean spectral density is given by
\be\label{spctrl_density_large_freq}
\langle S(\omega)\rangle_{\rm ME} \simeq  \frac{\pi\hbar}{a_V \langle Z(\beta)\rangle_{\rm ME}} \int dE \, {\rm e}^{-\beta E} \, \sigma(E)\, \sigma(E-\hbar\omega) \, .
\ee
Since the average density of states has the semicircular form~(\ref{semicircle_law}), the integral over energy is not equal to zero as long as there is an overlap between the two semicircular densities $\sigma(E)$ and $\sigma(E-\hbar\omega)$, i.e., if the frequency belongs to the interval $-2 E_{\rm b} < \hbar\omega < +2 E_{\rm b}$, where $E_{\rm b}=\sqrt{\frac{2N}{a_{H_0}}}$ is the half-width of the energy band.  The mean spectral density is equal to zero for frequencies outside this interval.
\\

At the edges of this interval, the mean spectral density vanishes in a way that can be calculated.  Indeed, the overlap between the two semicircular densities shrinks to zero as $\hbar\omega \to \pm 2 E_{\rm B}$, so that the integral extends over a decreasing energy interval.  As a consequence, the mean spectral density vanishes at its edges according to
\be\label{spctrl_density_edges}
\hbar\omega \to \pm 2 E_{\rm B}: \qquad 
\langle S(\omega)\rangle_{\rm ME} \simeq \frac{\hbar(g_\nu a_{H_0})^2}{a_V \langle Z(\beta)\rangle_{\rm ME}} \, E_{\rm b} \, {\rm e}^{\mp\beta E_{\rm b}} \left(\frac{\hbar\omega}{2}\mp E_{\rm b}\right)^2  ,
\ee
as shown in Appendix~\ref{app:spctrl_density}.  This behaviour is observed in Figs.~\ref{Fig9} and~\ref{Fig10}, where the edges are located at $\pm 2 E_{\rm b}=\pm 28.3$, because $N=100$ and $a_{H_0}=1$.

\subsubsection{Small-frequency behaviour}

For frequencies smaller than the mean spacing, i.e., for $\vert\hbar\omega\vert \lesssim \langle E_{m+1}-E_m\rangle_{\rm ME}$, the two-point correlation function $Y(\Delta e)$ is no longer vanishing, but given by Eq.~(\ref{Y-small}).  Inserting this behaviour into Eq.~(\ref{Mspctrl-Y}), using Eq.~(\ref{De-omega}), and expanding at small frequencies, we obtain
\be\label{Mspctrl-Y-small}
\langle S(\omega)\rangle_{\rm ME} \simeq \frac{\pi}{a_V} \, (\gamma_\nu + g_\nu) \, \delta(\omega) + \frac{\pi\hbar}{a_V} \, A_\nu \, \vert\hbar\omega\vert^{\nu} + O\Big( \vert\omega\vert^{\nu + 1}\Big)
\ee
with the same coefficient $A_\nu$ as the one given by Eq.~(\ref{coeff-A_nu}), because the mean spectral density is the Fourier transform of the mean two-time correlation function, which has the algebraic power-law decay~(\ref{Mcorrel-long-time}) at long times.  The result~(\ref{Mspctrl-Y-small}) shows that the spectral density presents not only the delta peak at zero frequency, but also a dip, which finds its origin in the universal level repulsion and which extends over a tiny frequency range limited by the mean spacing between the energy levels, $\vert\omega\vert \lesssim \hbar^{-1}\langle E_{m+1}-E_m\rangle_{\rm ME}$.  This dip is observed in Fig.~\ref{Fig10}(b) and it scales to zero with the expected universal exponent, as confirmed in Fig.~\ref{Fig10}(c) plotting the cumulative function defined by
\be\label{F(w)}
F(\omega) \equiv \int_{0^+}^{\omega} \langle S(\omega')\rangle_{\rm ME} \, d\omega' \, ,
\ee
which scales as $F(\omega)\sim \omega^{\nu+1}$ because of Eq.~(\ref{Mspctrl-Y-small}).  We note that the dip is also present around $\omega=0$ in Fig.~\ref{Fig9}, but less clearly visible because the cells of the histogram are therein ten times broader than in Fig.~\ref{Fig10}.

%%%%%%%%%%%%%%%%%%%%%%%%%%%%%%%%%%%%%%%%%%%%%%%%%
\section{Higher-order response functions}
\label{sec:higher-resp}

\subsection{The third-order response and four-time correlation functions}

The four-time correlation function is defined as
\be\label{dfn-4correl}
C(t_1,t_2,t_3) = {\rm tr}\, \hat\rho_0 \, \hat V(0) \, \hat V(t_1) \, \hat V(t_2) \, \hat V(t_3) = {\rm tr}\, \hat\rho_0 \, \hat V(\tau) \, \hat V(\tau+t_1) \, \hat V(\tau+t_2) \, \hat V(\tau+t_3)
\ee
in terms of the canonical statistical operator~(\ref{rho0}) and the time-evolved perturbation operator~(\ref{V(t)}).  Like the two-time correlation function~(\ref{dfn-correl}), it is invariant under time translations and, thus, it only depends on three consecutive times.
Using the notation $\hat U(t)=\exp(-{\rm i}\hat H_0t/\hbar)$ for the unitary evolution operator, the correlation function also reads
\be\label{4correl-fn}
C(t_1,t_2,t_3) = \frac{1}{Z(\beta)} \, {\rm tr}\left[ \hat U(t_3-{\rm i}\beta\hbar) \, \hat V \, \hat U(-t_1) \, \hat V \, \hat U(t_1-t_2) \, \hat V \, \hat U(t_2-t_3) \, \hat V \right]
\ee
with the partition function~(\ref{Z}).  Since the trace is invariant under cyclic permutations of the operators, the following relations hold,
\bea\label{sym_rel_4correl}
C(t_1,t_2,t_3) &=& C(t_2-t_1,t_3-t_1,-t_1+{\rm i}\beta\hbar) = C(t_3-t_2,-t_2+{\rm i}\beta\hbar,t_1-t_2+{\rm i}\beta\hbar) \nonumber\\
&=& C(-t_3+{\rm i}\beta\hbar,t_1-t_3+{\rm i}\beta\hbar,t_2-t_3+{\rm i}\beta\hbar) \, .
\eea
Furthermore, its complex conjugate is given by
\be\label{cc_4correl}
C^*(t_1,t_2,t_3) = C(t_2-t_3,t_1-t_3,-t_3) \, .
\ee
In addition, the four-time correlation can be decomposed as
\be\label{4correl-fn-bis}
C(t_1,t_2,t_3) = \frac{1}{Z(\beta)} \, \sum_{k,l,m,n} \left[ U_k(t_3-{\rm i}\beta\hbar) \, V_{kl} \, U_l(-t_1) \, V_{lm} \, U_m(t_1-t_2) \, V_{mn} \, U_n(t_2-t_3) \, V_{nk} \right]
\ee
onto the orthonormal basis of eigenstates of the Hamiltonian operator, $\hat H_0\vert k \rangle = E_k\vert k\rangle$, where
\be
\langle k \vert \hat U(t) \vert l \rangle = \delta_{kl} \, U_k(t) = \delta_{kl} \, {\rm e}^{-\frac{\rm i}{\hbar} E_k \, t} \, .
\ee
\\

Now, the third-order mean response function is given by Eq.~(\ref{MRFn}) with $n=3$, which can be written as
\bea
R_{VVVV}(t-t_1,t-t_2,t-t_3) &=& \frac{1}{({\rm i}\hbar)^3}\, \Big\langle {\rm tr} \, \hat\rho_{0} \Big[\Big[\Big[ \hat V(t), \hat V(t_1) \Big],\hat V(t_2) \Big],\hat V(t_3)\Big] \Big\rangle_{\rm ME} \nonumber\\
&=& -\frac{2}{\hbar^2}\, {\rm Im} \Big\langle {\rm tr} \, \hat\rho_{0} \Big[\Big[ \hat V(t), \hat V(t_1) \Big],\hat V(t_2) \Big]\hat V(t_3) \Big\rangle_{\rm ME} \, .
\label{MRF3}
\eea
Setting $\tau_j=t-t_j$, this mean response function can be expressed as
\bea
R_{VVVV}(\tau_1,\tau_2,\tau_3) &=& -\frac{2}{\hbar^2}\, {\rm Im} \Big\langle C(-\tau_1,-\tau_2,-\tau_3) - C(\tau_1,\tau_1-\tau_2,\tau_1-\tau_3) \nonumber\\
&&\qquad\qquad - C(\tau_2,\tau_2-\tau_1,\tau_2-\tau_3) + C(\tau_2-\tau_1,\tau_2,\tau_2-\tau_3)\Big\rangle_{\rm ME}
\label{MRF3-bis}
\eea
in terms of the four-time correlation function~(\ref{dfn-4correl}).
\\

The next issue is to evaluate the statistical average over the matrix ensemble of the four-time correlation function~(\ref{4correl-fn-bis}), which is given by
\be\label{M-4correl-fn}
\langle C(t_1,t_2,t_3)\rangle_{\rm ME}  = \sum_{k,l,m,n} \left\langle\frac{1}{Z(\beta)} \, U_k(t_3-{\rm i}\beta\hbar) \, U_l(-t_1) \, U_m(t_1-t_2) \, U_n(t_2-t_3) \right\rangle_{\rm ME} \langle V_{kl} V_{lm} V_{mn} V_{nk} \rangle_{\rm ME} \, .
\ee
The Wick lemma can be used in order to evaluate the statistical averages of products of matrix elements of the perturbation operator $\hat V$ over the Gaussian probability distribution~(\ref{P(V)}).  In particular, the product of four matrix elements can be expanded as follows in terms of the corresponding covariances,
\be
\langle V_{kl} V_{lm} V_{mn} V_{nk} \rangle_{\rm ME} = \langle V_{kl} V_{lm}\rangle_{\rm ME} \, \langle V_{mn} V_{nk} \rangle_{\rm ME} + \langle V_{kl} V_{mn}\rangle_{\rm ME} \, \langle V_{lm} V_{nk} \rangle_{\rm ME} + \langle V_{kl} V_{nk}\rangle_{\rm ME} \, \langle V_{lm} V_{mn} \rangle_{\rm ME} 
\ee
and the products of odd numbers of matrix elements are equal to zero.  In the GOE, the covariances of the matrix elements are given by
\be
\langle V_{kl} V_{mn}\rangle_{\rm GOE} =\frac{1}{2 a_V} \left( \delta_{km}\, \delta_{ln} + \delta_{kn}\, \delta_{lm}\right) ,
\ee
which reduces to Eq.~(\ref{var-V}) with $\nu=1$, if $k=m$ and $l=n$, as expected.  Accordingly, we obtain
\bea
\langle V_{kl} V_{lm} V_{mn} V_{nk} \rangle_{\rm GOE} &=& \frac{1}{4 a_V^2} \Big( 5 \ \delta_{kl}\, \delta_{lm} \, \delta_{mn}\, \delta_{nk} \nonumber\\
&&\qquad +\;  \delta_{kl}\, \delta_{lm} \, \delta_{mk} + \delta_{km}\, \delta_{mn} \, \delta_{nk} + \delta_{kl} \, \delta_{ln} \, \delta_{nk}+  \delta_{lm} \, \delta_{mn} \, \delta_{nl}+ \delta_{km}\, \delta_{ln} + \delta_{km} + \delta_{ln} \Big) \, .
\label{Vs-deltas}
\eea
Inserting this expression into Eq.~(\ref{M-4correl-fn}) and using the contraction rule $\delta_{mm'} U_m({\mathscr T}) U_{m'}({\mathscr T}') = \delta_{mm'} U_m({\mathscr T}+{\mathscr T}')$, we find in the GOE case that the intermediate-time behaviour of the mean four-time correlation function is given by 
\bea\label{M-4correl-fn-approx}
\langle C(t_1,t_2,t_3)\rangle_{\rm GOE}  \simeq \frac{1}{4 a_V^2 {\mathscr Z}(\beta)} \biggl\{ 5 \,  {\mathscr Z}(\beta) 
+  {\mathscr Z}\Big[\beta-\frac{\rm i}{\hbar}(t_2-t_3)\Big] \,  {\mathscr Z}\Big[\frac{\rm i}{\hbar}(t_2-t_3)\Big]
+  {\mathscr Z}\Big(\beta+\frac{\rm i}{\hbar}t_1\Big) \,  {\mathscr Z}\Big(-\frac{\rm i}{\hbar}t_1\Big) \nonumber\\
+  {\mathscr Z}\Big[\beta-\frac{\rm i}{\hbar}(t_1-t_2)\Big] \,  {\mathscr Z}\Big[\frac{\rm i}{\hbar}(t_1-t_2)\Big]
+  {\mathscr Z}\Big(\beta+\frac{\rm i}{\hbar}t_3\Big) \,  {\mathscr Z}\Big(-\frac{\rm i}{\hbar}t_3\Big) \nonumber\\
+  {\mathscr Z}\Big[\beta+\frac{\rm i}{\hbar}(t_1-t_2+t_3)\Big] \,  {\mathscr Z}\Big[-\frac{\rm i}{\hbar}(t_1-t_2+t_3)\Big] \nonumber\\
+  {\mathscr Z}\Big[\beta+\frac{\rm i}{\hbar}(t_1-t_2+t_3)\Big] \,  {\mathscr Z}\Big(-\frac{\rm i}{\hbar}t_1\Big) \,  {\mathscr Z}\Big[\frac{\rm i}{\hbar}(t_2-t_3)\Big] \nonumber\\
+ {\mathscr Z}\Big(\beta+\frac{\rm i}{\hbar}t_3\Big) \,  {\mathscr Z}\Big[-\frac{\rm i}{\hbar}(t_1-t_2+t_3)\Big] \, {\mathscr Z}\Big[\frac{\rm i}{\hbar}(t_1-t_2)\Big]
\biggr\}
\eea
in terms of the functions~(\ref{Z(b)}), (\ref{Z(b+it)}), and~(\ref{Z(it)}) with $\nu=1$.  As a consequence, we have in particular that
\be\label{M-4correl-fn-approx-0}
\langle C(0,0,0)\rangle_{\rm GOE}  = \frac{1}{4 a_V^2} \left( 5+ 5N+2N^2\right) ,
\ee
because ${\mathscr Z}(0)=N$.  Furthermore, the mean correlation function~(\ref{M-4correl-fn-approx}) decays to zero in the limit of arbitrarily large separations between the successive times, i.e., for $\vert t_{j+1}-t_j\vert\to\infty$ with $j=1,2$, although the precise decay law should be obtained in the long-time limit by keeping the matrix-ensemble average over the products, as we have seen in the previous section for the mean two-time correlation function.

\subsection{The diagrammatic method and multiple-time correlation functions}

By generalisation, we see that the $n^{\rm th}$-order response function~(\ref{MRFn}) can be expressed in terms of ($n+1$)-time correlation functions defined as
\be\label{dfn-(n+1)correl}
C(t_1,t_2,\dots,t_n) = {\rm tr}\, \hat\rho_0 \, \hat V(0) \, \hat V(t_1) \, \hat V(t_2) \, \cdots \, \hat V(t_n) \, .
\ee
By analogy with Eq.~(\ref{M-4correl-fn}), its mean value is given by
\bea\label{M-(n+1)correl-fn}
\langle C(t_1,t_2,\dots,t_n)\rangle_{\rm ME}  = \sum_{k_1,\dots,k_{n+1}} \left\langle\frac{1}{Z(\beta)} \, U_{k_{n+1}}(t_n-{\rm i}\beta\hbar) \, U_{k_1}(-t_1) \, U_{k_2}(t_1-t_2) \, \cdots \, U_{k_n}(t_{n-1}-t_n) \right\rangle_{\rm ME} \nonumber\\
\times \langle V_{k_{n+1}k_1} V_{k_1k_2} V_{k_2k_3} \cdots V_{k_nk_{n+1}} \rangle_{\rm ME} \, .
\eea
For Gaussian random-matrix ensembles, the statistical average over the product of matrix elements $V_{k_j k_{j+1}}$ can be evaluated using the Wick lemma and formulas such as Eq.~(\ref{Vs-deltas}). The contributions of the mean values of the matrix elements $V_{k_jk_{j+1}}$ are shown in Table~\ref{TableI}.  The result can be represented by diagrams composed of vertices and links.  The vertices of these diagrams have the labels of the eigenstates $k_j$.  Each vertex is associated with a sum $\sum_{n} U_{n}(\sum_i{\mathscr T}_i)$ obtained with the contraction rule $\delta_{mm'} U_m({\mathscr T}) U_{m'}({\mathscr T}') = \delta_{mm'} U_m({\mathscr T}+{\mathscr T}')$ applied to all the factors $U_{k_j}({\mathscr T})$ with the same label $k_j$.  Each link is associated with a matrix element $V_{k_j k_{j+1}}$.  There are $n+1$ links in all the diagrams contributing to the mean ($n+1$)-time correlation function.  Before summation over the labels $\{k_j\}$, the vertices are consecutively linked by lines in the order $k_1,k_2,k_3,\dots,k_{n+1}$ and they form a loop, $k_{n+1}$ being finally linked to $k_1$.
\\
%%%%%%%%%%%%%%%%%%%%%%%%%%%%%%%%%%%%%%%%%%%%%%%%%%%%%%%%%%
\begin{table}[h!]
{\includegraphics[width=0.7\textwidth]{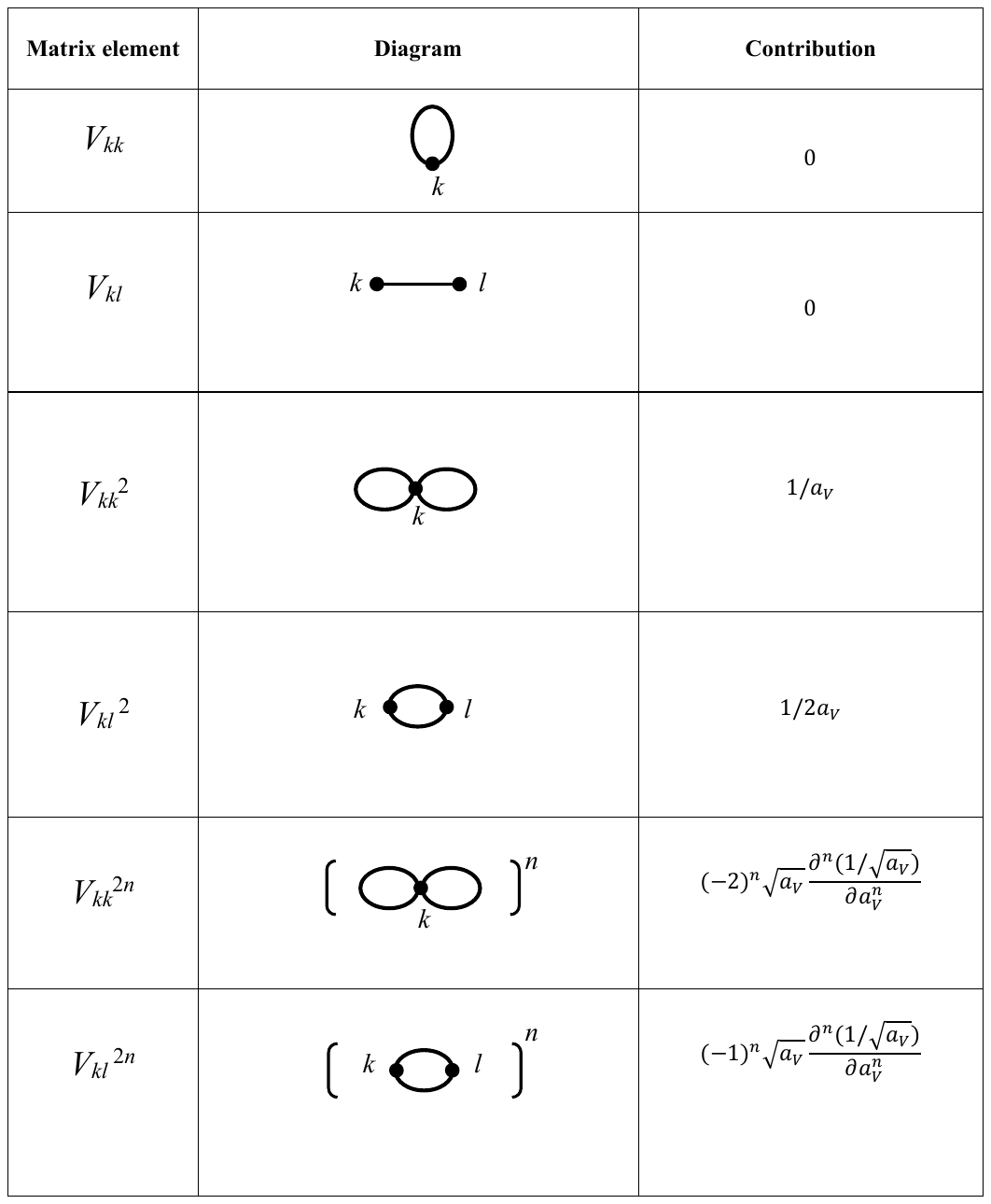}}
\caption[] {Basic diagrams of the matrix elements and their contributions, in terms of which the averages of any product of matrix elements appearing in a Gaussian random-matrix calculation can be reduced.}
\label{TableI}
\end{table}
%%%%%%%%%%%%%%%%%%%%%%%%%%%%%%%%%%%%%%%%%%%%%%%%%%%%%%%%%%
\begin{table}[h!]
{\includegraphics[width=0.7\textwidth]{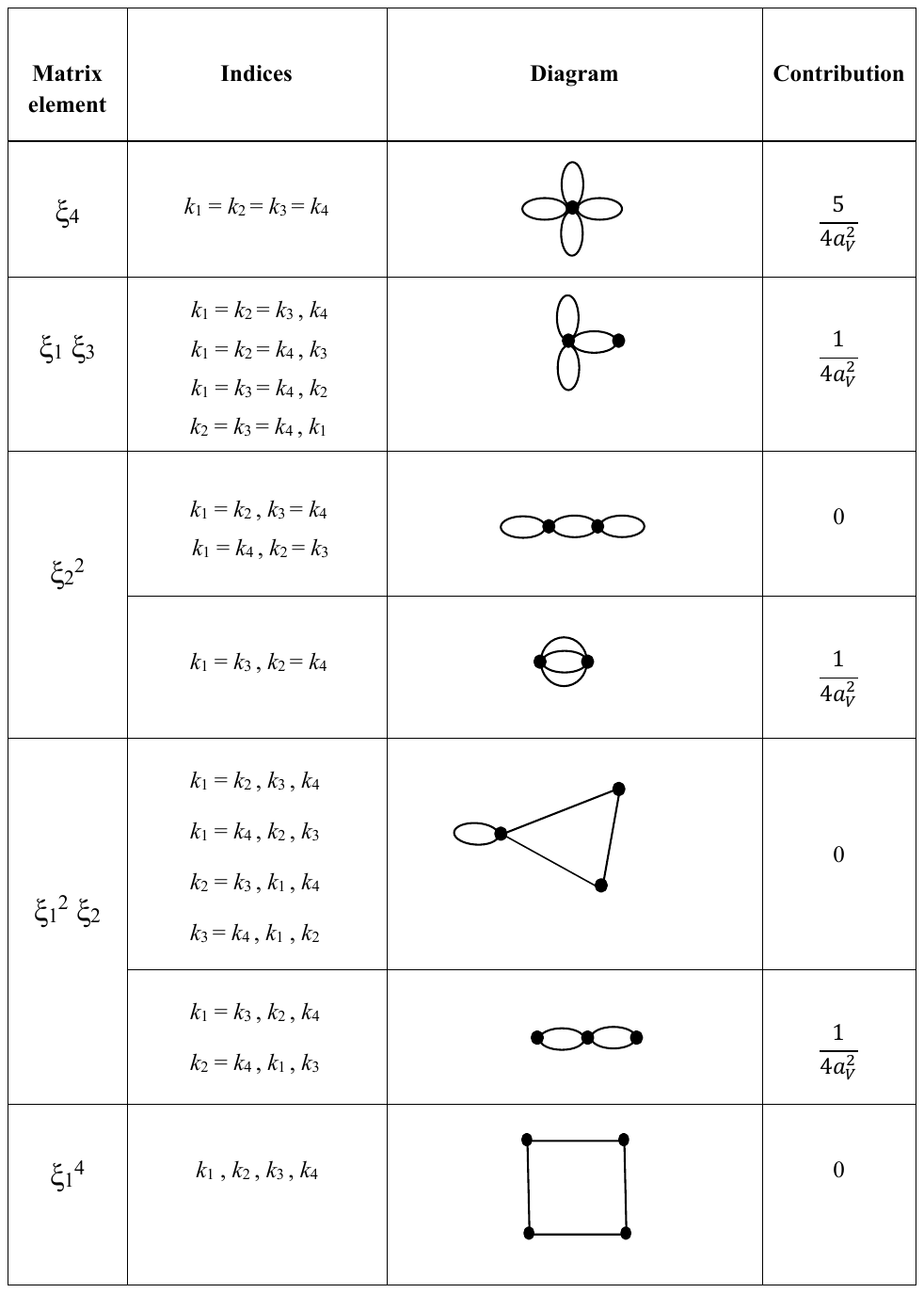}}
\caption[] {The polynomial~(\ref{Polynomial-A4}) of order $4$ appearing in the calculation of the $4$-time correlation function has terms appearing in the first column.  For each term, there are relations between indices which come from the ordered cycle indicator function shown in the second column.  Third and fourth columns then show the diagram and its contribution in accordance with Table~\ref{TableI}.}
\label{TableII}
\end{table}
%%%%%%%%%%%%%%%%%%%%%%%%%%%%%%%%%%%%%%%%%%%%%%%%%%%%%%%%%%
\begin{table}[h!]
{\includegraphics[width=0.7\textwidth]{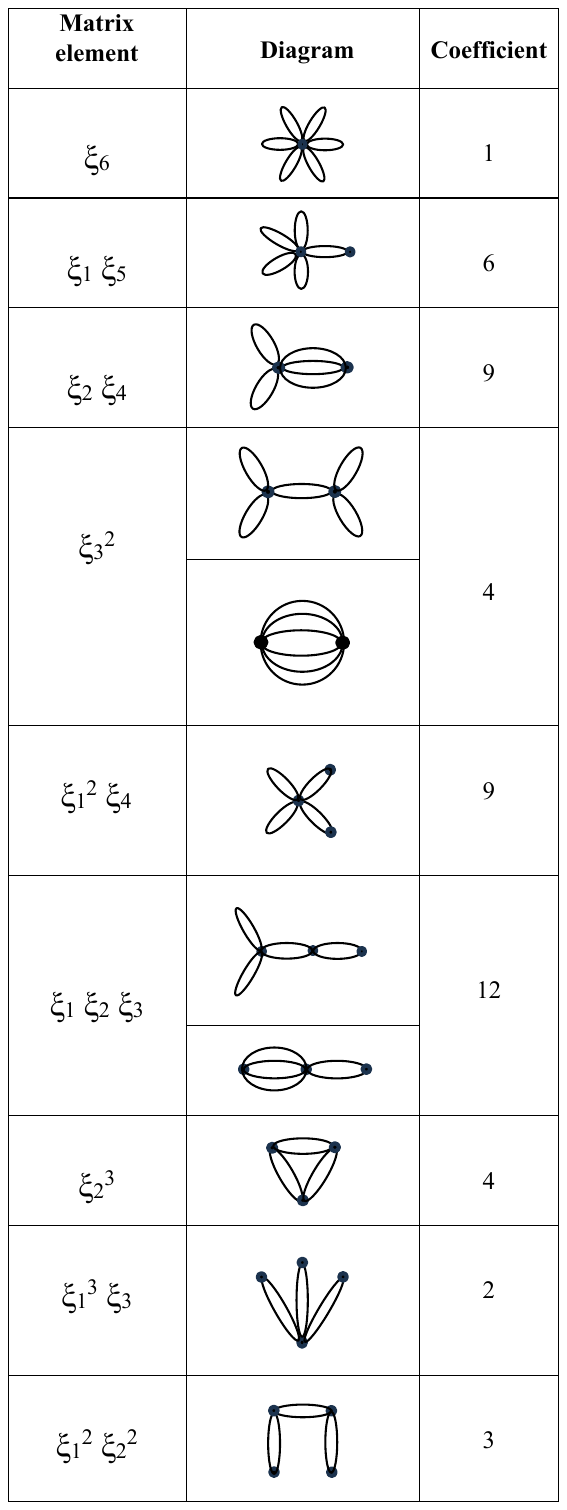}}
\caption[] {The polynomial~(\ref{Polynomial-R6}) of order $6$ appearing in the calculation of the $6$-time correlation function has terms appearing in the first column.  For each term, there are relations between indices which come from the ordered cycle indicator function, diagrams corresponding to these are shown in the second column.  Third column gives the coefficient of the terms.}
\label{TableIII}
\end{table}
%%%%%%%%%%%%%%%%%%%%%%%%%%%%%%%%%%%%%%%%%%%%%%%%%%%%%%%%%%

The different possible contributions can be enumerated with the Bell polynomials~\cite{B39} and the di~Bruno formula~\cite{R58}.  Indeed, the summation in Eq.~(\ref{M-(n+1)correl-fn}) can be carried out in the
following manner. We have $J=n+1$ indices. Different situations emerge when several
of them are equal or unequal. We can construct cycles of indices which are
equal, and within each cycle we do not need to count permutations. Thus we have
ordered cycles \cite{R58}. So, the $J$ indices decompose into classes of ordered
cycles of permutations. And the sum reduces to a sum over classes where in a
given class, there are all possible distinct permutations, i.e.,
\be
\sum_{k_1,\dots,k_J} = \sum_{\rm classes} \quad \sum_{{\cal
P}_{\rm classes}} \, ,
\ee
where ${\cal P}_{\rm classes}$ denotes permutations of indices within a class. 
The classes are labelled by non-negative integers $(m_1 m_2 \dots m_J)$, which satisfy
\be
1 \times m_1 + 2 \times m_2 + \cdots + J \times m_J = J\, . 
\label{constraint}
\ee
Let us explain in some detail when $J=n+1=4$, say. The possible classes are
$(m_1m_2m_3m_4) =$ $(0001)$, $(1010)$, $(0200)$, $(2100)$, $(4000)$. The class $(0001)$ consists of a 4-cluster
($k_1,k_2,k_3,k_4$) which means that $k_1=k_2=k_3=k_4$. Calling an $l$-cluster by $\xi _l$,
$b(\xi _4) = 1$ gives the number of 4-clusters, which obviously equals unity.
Similarly, the class $(1010)$ corresponds to one 1-cluster and one 3-cluster,
i.e., indices grouped as $(k_1,k_2,k_3)(k_4)$, $(k_1,k_2,k_4)(k_3)$, $(k_1,k_3,k_4)(k_2)$, $(k_2,k_3,k_4)(k_1)$ thus
$b(1010)=b(\xi _1 \xi _3) = 4$. Next, we have the class $(0200)$ with two 2-clusters of indices grouped as $(k_1,k_2)(k_3,k_4)$, $(k_1,k_3)(k_2,k_4)$, $(k_1k_4)(k_2,k_3)$, so that $b(0200)=b(\xi_2^2)=3$.  The class $(2100)$ has two 1-clusters and one 2-cluster such that $(k_1,k_2)(k_3)(k_4)$, $(k_1,k_3)(k_2)(k_4)$, $(k_1,k_4)(k_2)(k_3)$, $(k_2,k_3)(k_1)(k_4)$, $(k_2,k_4)(k_1)(k_3)$, $(k_3,k_4)(k_1)(k_2)$, giving $b(2100)=b(\xi_1^2\xi_2)=6$.  Finally, the class $(4000)$ contains four
1-clusters, $(k_1)(k_2)(k_3)(k_4)$ and $b(4000)=b(\xi _1^4)=1$. We have, consequently, a
polynomial,
\bea
{\cal B}_4 &=&  b(\xi _4 ) \, \xi _4  + b(\xi _1\xi _3)\, \xi _1\xi _3 + b(\xi _2^2)\, \xi _2^2 + b(\xi _1^2\xi _2)\, \xi _1^2\xi _2 + b(\xi _1^4)\, \xi _1^4\nonumber \\  
&=& \xi _4 + 4\, \xi _1\xi _3  + 3\, \xi _2^2  + 6\, \xi _1^2\xi _2 + \xi _1^4 \, .
\label{Polynomial-A4}
\eea
This is one of the ordered cycle indicator function of the permutation
group, and is an example of Bell polynomials~\cite{B39}.  In general, when there are $J=n+1$
indices, the indicator function is given by the di Bruno formula \cite{R58}
\begin{equation}
{\cal B}_{J} = \sum_{\{m_j\}} \frac{J!}{m_1! \cdots m_J!}\left(\frac{{\xi _1} }{
{1!}}\right)^{m_1} \cdots \left(\frac{{\xi _J} }{ {J!}}\right)^{m_J} \, ,
\end{equation}
where the summation is subjected to the constraint~(\ref{constraint}). With
these general considerations, one can enumerate the diagrams contributing to the statistical average of a multiple-time
correlation function.  In particular, the total number of possible diagrams is given by ${\cal B}_J$ with $\xi_1=\cdots=\xi_J=1$.
\\

The diagrams contributing to the mean four-time correlation function are shown in Table~\ref{TableII}.  
The diagrams corresponding to each of the terms of the polynomial~(\ref{Polynomial-A4}), starting with the largest
cluster size $\xi _4$, are given in Table~\ref{TableII}. The contribution of different
diagrams can be easily written down by using Table~\ref{TableI}. However, as can be seen
already from Table~\ref{TableII}, not all the diagrams are non-zero in each term. In consistency with Table~\ref{TableI}, following rules may be enunciated: (i) for a non-zero contributing term, each pair of vertices must share a pair (pairs) of link(s); (ii) an isolated vertex must be self-connected with an even number of links; (iii) a diagram is a connected graph. It would be useful to be able to write a set of polynomials with non-zero contributing terms only. The first non-trivial polynomial can be written by referring to Table~\ref{TableII}, denoting it by ${\mathcal R}_4$:
\bea
{\mathcal R}_4 &=& \xi _4 + 4 \, \xi _1 \xi _2 + \xi _2^2 +  2 \, \xi _1^2 \xi _2 \, .
\label{Polynomial-R4}
\eea
The degree of each term in the polynomial gives us the number of vertices. For an $n$-time correlation function, the maximum number of vertices for non-zero contributing diagrams is $(n/2 + 1)$.  

The method extends to higher-order mean correlation functions. For instance, for the six-time correlation function, the relevant polynomial is 
\bea
{\mathcal R}_6 &=& \xi _6 + 6 \, \xi _1 \xi _5 + 9 \, \xi _2 \xi _4 + 4 \, \xi _3^2 + 9 \, \xi _1^2 \xi _4 + 12 \, \xi _1 \xi _2\xi _3 + 4 \, \xi _2^3 + 2 \, \xi _1^3 \xi _3 + 3 \, \xi _1 ^2 \xi _2^2 \, . 
\label{Polynomial-R6}
\eea
Table~\ref{TableIII} shows all the non-zero diagrams. 

The combinatorial problem becomes more complex for higher-order correlation functions. From all possible  diagrams for the eight-time correlation function, the non-zero diagrams can be summarized into the following polynomial:
\bea
{\mathcal R}_8 &=& \xi _8 + 8 \, \xi _1 \xi _7 + 20 \, \xi _2 \xi _6 + 20 \, \xi _1^2 \xi _6 + 24 \, \xi _3 \xi _5 + 56 \, \xi _1 \xi _2 \xi _5 + 16 \, \xi _1^3 \xi _5 + 19 \, \xi _4^2 + 56 \, \xi _1 \xi _3 \xi _4 \nonumber \\ &&+ 54 \, \xi _2^2 \xi _4 + 36 \, \xi _1^2  \xi _2 \xi _4 + 2 \, \xi _1^4 \xi _4 + 40 \, \xi _2 \xi _3^2 + 16 \, \xi _1^2 \xi _3^2 + 48 \, \xi _1 \xi _2^2 \xi _3 + 8 \, \xi _1^3 \xi _2 \xi _3 + 5 \, \xi _2^4 + 4 \, \xi _1^2 \xi _2^3 \, . 
\label{Polynomial-R8}
\eea 
Symbolic programs can be employed to obtain these new class of polynomials, which are metaphorically related to connecting a given number of points with an even pairs of links between each pair, without lifting the hand.   

%%%%%%%%%%%%%%%%%%%%%%%%%%%%%%%%%%%%%%%%%%%%%%%%%%%%%%%%%%
%%%%%%%%%%%%%%%%%%%%%%%%%%%%%%%%%%%%%%%%%%%%%%%%%
\section{Dynamical response to impulsive perturbations}
\label{sec:impulsive}

\subsection{Pulse train}

Here, we apply the results of the previous sections to the response of the system to a pulse train given by
\be
\lambda(t) = \sum_{\mu=1}^{M} p_{\mu}(t-{\mathscr T}_{\mu})
\qquad\mbox{with}\qquad
p_{\mu}(\Delta t) = \frac{\alpha_\mu}{\sqrt{2\pi w_\mu^2}} \, \exp\left(-\frac{\Delta t^2}{2 w_\mu^2}\right) ,
\ee
where the pulses have the amplitudes $\alpha_\mu$, the widths $w_\mu$, and occur at the successive times ${\mathscr T}_{\mu}$.  The pulses are assumed to have a width much smaller than the characteristic time of the response, $w_\mu \ll \tau_{\rm b}$, and to be devoid of oscillations.  Moreover, the separations between the successive pulse times is supposed to be significantly longer than the response characteristic time, $\tau_{\rm b} \ll \vert{\mathscr T}_{\mu +1}-{\mathscr T}_{\mu}\vert$ for all the pulses $\mu$.  In this limit of ultrashort pulses, the perturbation is said to be impulsive.

\subsection{Response at first and third orders}

The response~(\ref{response-A}) of the system for the mean value of the perturbation operator itself~(\ref{A=V}) can be divided by the number $P$ of parts composing the system in the limit $P\to\infty$, so that we obtain its statistical average over the random-matrix ensemble according to
\be
\langle\langle \hat V \rangle_t\rangle_{\rm ME} = \langle\langle \hat V \rangle_t^{(1)}\rangle_{\rm ME} + \langle\langle \hat V \rangle_t^{(3)}\rangle_{\rm ME} + O(\lambda^5) \, ,
\ee
where
\bea
\langle\langle \hat V \rangle_t^{(1)}\rangle_{\rm ME} &=& \int_{-\infty}^{+\infty} dt_1 \, \lambda(t_1) \, \theta(t-t_1) \, R_{VV}(t-t_1) \, , \label{MV1} \\
\langle\langle \hat V \rangle_t^{(3)}\rangle_{\rm ME} &=& \int_{-\infty}^{+\infty} dt_1 \int_{-\infty}^{+\infty} dt_2 \int_{-\infty}^{+\infty}  dt_3 \, \lambda(t_1) \, \lambda(t_2) \, \lambda(t_3) \, \theta(t-t_1) \, \theta(t_1-t_2) \, \theta(t_2-t_3) \, R_{VVVV}(t-t_1,t-t_2,t-t_3) \, , \nonumber\\
\label{MV3}
\eea
in terms of the mean first- and third-order response functions~(\ref{MRF1}) and~(\ref{MRF3}).  There are no terms of even order,  because the ensembles are Gaussian and centered, so that the statistical averages with an odd power of the operator $\hat V$ are equal to zero.  In Eqs.~(\ref{MV1}) and~(\ref{MV3}), we have inserted Heaviside functions in order to satisfy the causality conditions $t_3<t_2<t_1<t$ and to extend to infinity the integrals over the different times.
\\

%%%%%%%%%%%%%%%%%%%%%%%%%%%%%%%%%%%%%%%%%%%%%%%%%%%%%%%%%%
\begin{figure}[h!]\centering
{\includegraphics[width=0.6\textwidth]{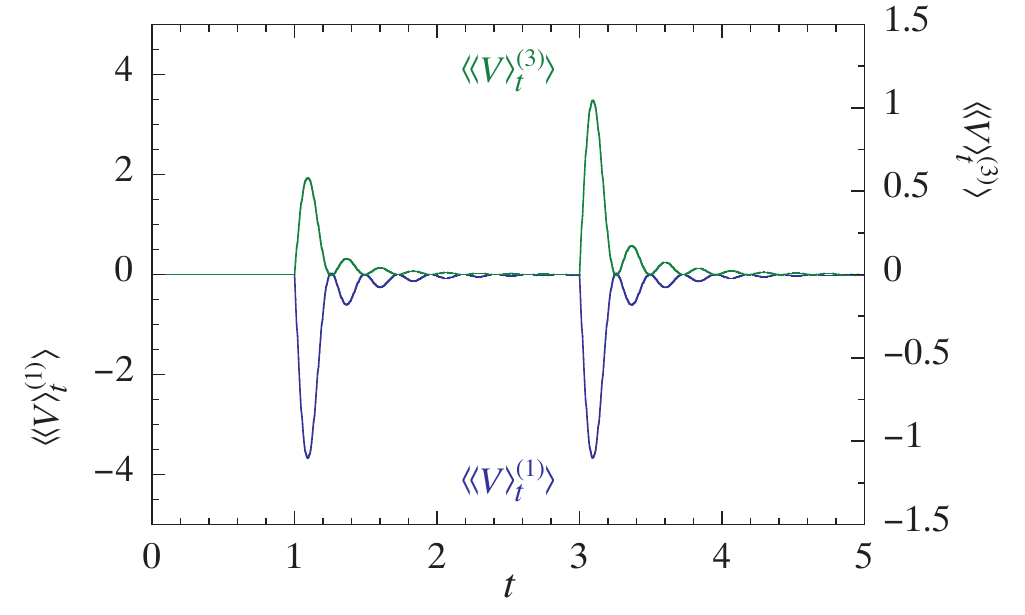}}
\caption[] {The contributions~(\ref{MV1-pulses}) and~(\ref{MV3-pulses}) to the dynamical response to impulsive perturbations by two pulses happening at times ${\mathscr T}_1=1$ and ${\mathscr T}_2=3$ with amplitudes $\alpha_1=\alpha_2=0.05$ for a quantum system of GOE $100\times 100$ matrices with $a_{H_0}=a_V=1$ at inverse temperature $\beta=1$.  These contributions are obtained using the intermediate-time approximations~(\ref{Mcorrel-Z-approx}) and~(\ref{M-4correl-fn-approx}) in terms of Bessel functions.}
\label{Fig11}
\end{figure}
%%%%%%%%%%%%%%%%%%%%%%%%%%%%%%%%%%%%%%%%%%%%%%%%%%%%%%%%%%

In the aforementioned limit of ultrashort pulses, the mean first-order response can be written as the following approximation,
\be
\langle\langle \hat V \rangle_t^{(1)}\rangle_{\rm ME} \simeq \sum_{\mu=1}^{M} \alpha_\mu \, \theta(t-{\mathscr T}_\mu) \, R_{VV}(t-{\mathscr T}_\mu) \, . 
\label{MV1-pulses}
\ee
However, in the mean third-order response, there is the possibility that a pulse has an overlap with itself, which can be taken into account by using the value $\theta(0)=1/2$ for the Heaviside function.  If we consider a train of $M=2$ pulses, e.g., in pump-probe spectroscopy~\cite{M95}, with the pump pulse at time $t={\mathscr T}_1$ and the probe pulse at time $t={\mathscr T}_2$, the mean third-order response can thus be approximated by
\bea
\langle\langle \hat V \rangle_t^{(3)}\rangle_{\rm ME} &=&\ \ \, \frac{1}{4} \, \alpha_1^3 \, \theta(t-{\mathscr T}_1) \, R_{VVVV}(t-{\mathscr T}_1,t-{\mathscr T}_1,t-{\mathscr T}_1) \nonumber\\
&&+\, \frac{1}{2} \, \alpha_1^2 \, \alpha_2 \, \theta(t-{\mathscr T}_2) \, R_{VVVV}(t-{\mathscr T}_2,t-{\mathscr T}_1,t-{\mathscr T}_1) \nonumber\\
&&+\, \frac{1}{2} \, \alpha_1 \, \alpha_2^2 \, \theta(t-{\mathscr T}_2) \, R_{VVVV}(t-{\mathscr T}_2,t-{\mathscr T}_2,t-{\mathscr T}_1)  \nonumber\\
&&+\, \frac{1}{4} \, \alpha_2^3 \, \theta(t-{\mathscr T}_2) \, R_{VVVV}(t-{\mathscr T}_2,t-{\mathscr T}_2,t-{\mathscr T}_2) \, .
\label{MV3-pulses}
\eea
\\
These results are illustrated in Fig.~\ref{Fig11}, showing the dynamical response to an impulsive perturbation composed of two pulses.  We see in Fig.~\ref{Fig11} that the contributions at first and third orders are qualitatively similar, although quantitatively different.

%%%%%%%%%%%%%%%%%%%%%%%%%%%%%%%%%%%%%%%%%%%%%%%%%
\section{Quantum fluctuations of individual members of the ensemble}
\label{sec:fluctuations}

\subsection{Probability distributions of the quantum fluctuations}

Until now, we have been dealing with ensemble-averaged correlation functions.  In this section, our focus is on an individual member or realisation participating in an ensemble.  For each such member in the GOE and GUE ensembles, the Hamiltonian operator $\hat H_0$ and the perturbation operator $\hat V$ are $N\times N$ matrices. There are thus $N$~eigenvalues for $\hat H_0$, so that the real and imaginary parts of the correlation function is given by the combinations~(\ref{Re-Im-correl}) of oscillating cosine and sine functions.  Such functions belong to the genera of almost-periodic functions~\cite{B23,B26}.
\\

\begin{figure}[h!]\centering
{\includegraphics[width=0.65\textwidth]{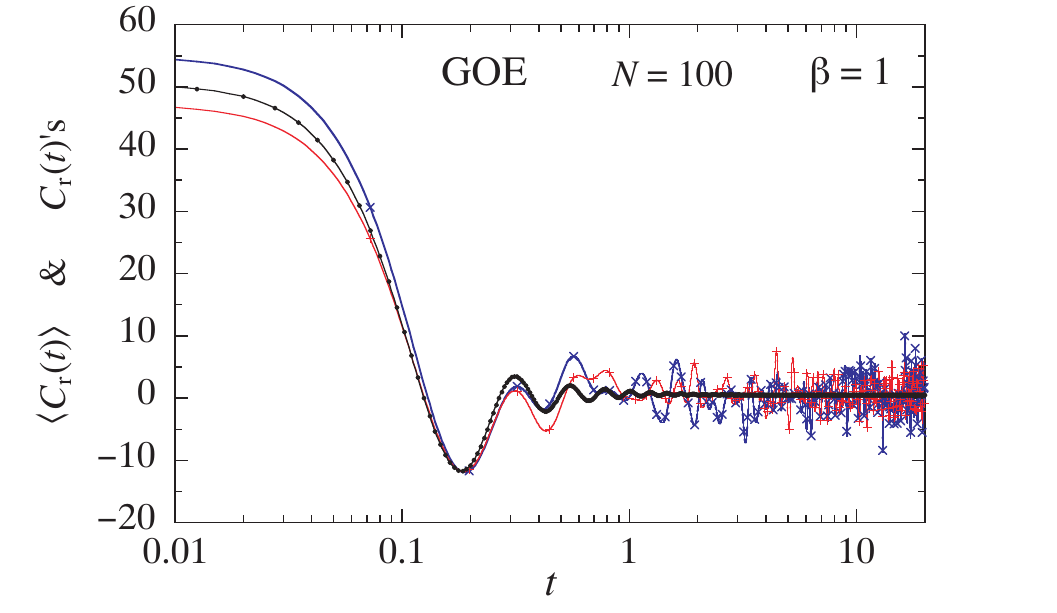}}
\caption[] {The real part of correlation functions~(\ref{correl-basis}) versus time for GOE $100\times 100$ matrices with $a_{H_0}=a_V=1$ at inverse temperature $\beta=1$.  The plot shows two random realisations (lines with crosses and pluses) and the mean value (line with circles) given by Eq.~(\ref{Mcorrel-Z-intermediate}) in terms of Bessel functions.}
\label{Fig12}
\end{figure}

\begin{figure}[h!]\centering
{\includegraphics[width=0.6\textwidth]{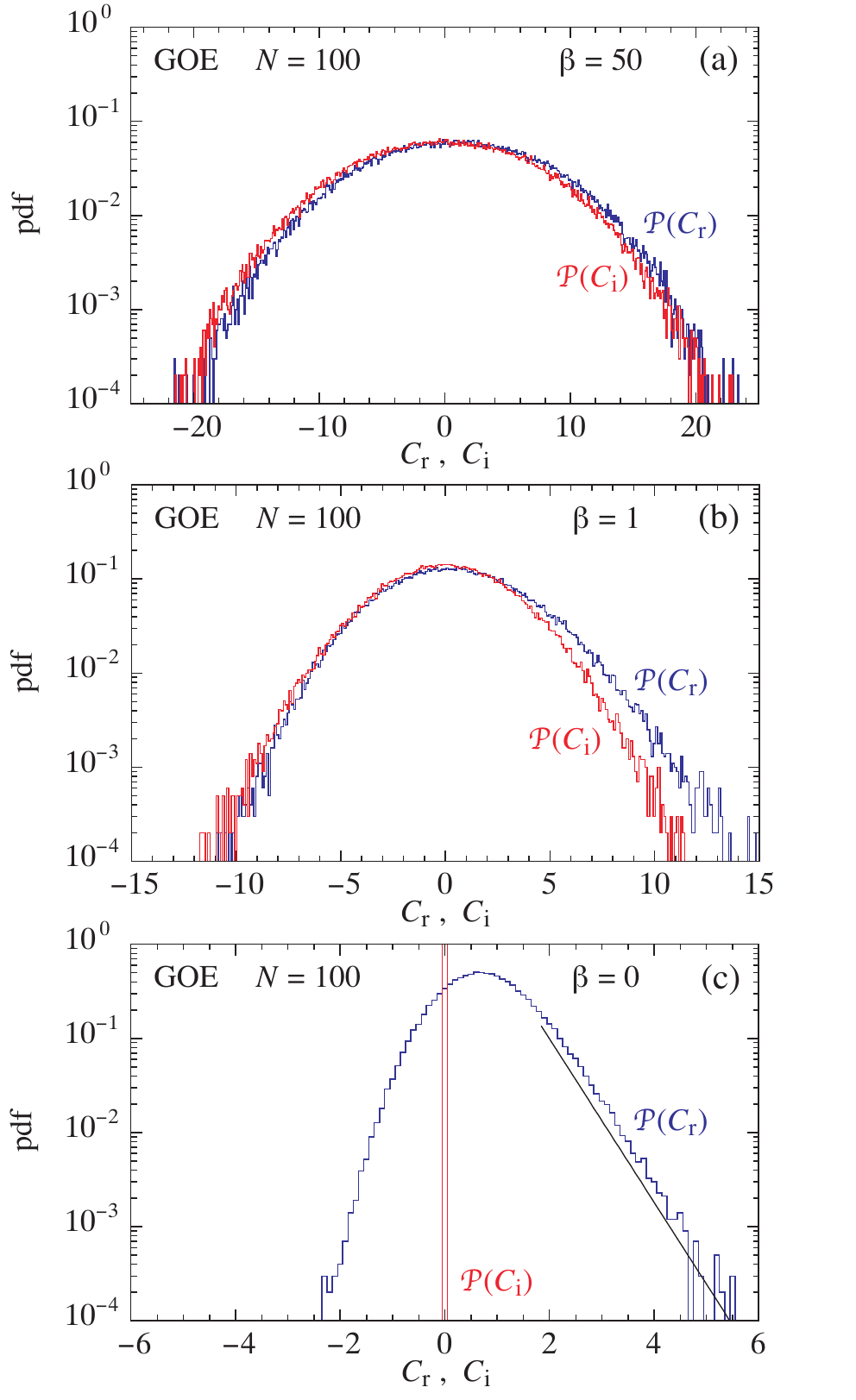}}
\caption[] {Probability densities~(\ref{dfn-P(C)}) for the real and imaginary parts of the two-time correlation functions for an individual member of the GOE system made of $100\times 100$ matrices with $a_{H_0}=a_V=1$ at inverse temperatures (a) $\beta=50$, (b) $\beta=1$, and (c) $\beta=0$.  Each function of time is sampled over $10^5$ time steps $\Delta t=1$ beyond time $t_0=10$.  The histograms are plotted for cells of size $\Delta C_{\rm r,i}=0.1$.  In panel~(c), the solid line shows the prediction of the exponential tail by Eq.~(\ref{exp-tail}).}
\label{Fig13}
\end{figure}

At early and intermediate times, before the oscillations start to dephase, the behaviour of the correlation function is well approximated by its ensemble average, if $N$ is large enough.  However, for times longer than the one corresponding to the the mean level spacing, i.e., for $t\gg \hbar/\sigma(E)$ at the mean thermal energy $E=\langle\hat H_0\rangle$, the correlation function manifests fluctuations due to the discrete spectrum for the individual members of the ensemble, as seen in Fig.~\ref{Fig12}.  These quantum fluctuations can be characterised in terms of the probability distribution of the real or imaginary part~(\ref{Re-Im-correl}) of the correlation function, i.e.,
\be\label{dfn-P(C)}
{\cal P}(C_{\rm r,i}) = \lim_{{\mathscr T}\to\infty} \frac{1}{\mathscr T} \int_0^{\mathscr T} \delta\big[ C_{\rm r,i} - C_{\rm r,i}(t)\big] \, dt
\ee
and their centered statistical moments.  Examples of such probability densities are shown in Fig.~\ref{Fig13}.
\\

We note that their first moment $\overline{C_{\rm r}(t)}=\overline{{\rm Re}\, C(t)}$ and $\overline{C_{\rm i}(t)}=\overline{{\rm Im}\, C(t)}$, i.e., the time averages themselves, are equal to the ensemble averages~(\ref{Re-Im-Mcorrel-infty}).  Their centered second moment or variance characterises the magnitude of the quantum fluctuations around their asymptotic mean value.  Depending on the temperature, these quantum fluctuations may have different kinds of probability distributions.  In the following, we focus on the GOE case to be specific.

%%%%%%%%%%%%%%%%%%%%%%%%%%%%%%%%%%%%%%%%%%%%%%%%%%%%%%%%%%
%%%%%%%%%%%%%%%%%%%%%%%%%%%%%%%%%%%%%%%%%%%%%%%%%%%%%%%%%%
%%%%%%%%%%%%%%%%%%%%%%%%%%%%%%%%%%%%%%%%%%%%%%%%%%%%%%%%%%
\begin{figure}[h!]\centering
{\includegraphics[width=0.5\textwidth]{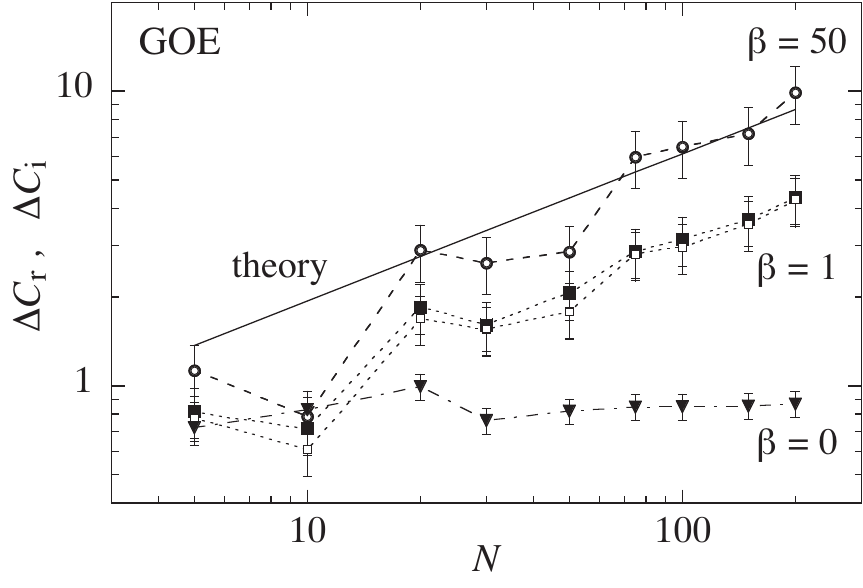}}
\caption[] {Log-log plot of the standard deviations $\Delta C_{\rm r,i}$ versus the matrix dimension $N$ for the long-time fluctuations of individual realisations of the real and imaginary parts of the correlation function for $N\times N$ random matrices with $a_{H_0}=a_V=1$ at inverse temperatures (a) $\beta=50$ (circles), (b) $\beta=1$ (squares), and (c) $\beta=0$ (triangles).  The solid line shows the theoretical predictions~(\ref{var-Cr-Ci}) at zero temperature, i.e., $\beta=\infty$. The filled symbols show the standard deviations $\Delta C_{\rm r}$ for the real parts and the open symbols the standard deviations $\Delta C_{\rm i}$ for the imaginary parts.}
\label{Fig14}
\end{figure}
%%%%%%%%%%%%%%%%%%%%%%%%%%%%%%%%%%%%%%%%%%%%%%%%%%%%%%%%%%

\subsection{The zero-temperature regime}

A simple case here is when $T=0$, i.e., $\beta=\infty$, so that the only positive Boltzmann probability weight is on the ground state denoted $\vert 0\rangle$ and corresponding to the minimum energy eigenvalue $E_0=E_{\rm min}$.  Hence, in the correlation function~(\ref{correl-basis}), we are left with a summation over one index and we have
\be\label{correl-basis-T=0}
C(t) = \sum_{l} \vert V_{0l}\vert^2 \, {\rm e}^{\frac{\rm i}{\hbar} (E_l-E_0)t} \, ,
\ee
which can be expressed as
\be\label{correl-basis-T=0-surv}
C(t) = {\rm e}^{-\frac{\rm i}{\hbar} E_0 t} \left\langle\psi\Big\vert {\rm e}^{\frac{\rm i}{\hbar}\hat H_0 t} \Big\vert \psi\right\rangle
\ee
in terms of the survival amplitude of the state $\vert\psi\rangle=\hat V\vert 0 \rangle$.  We note that the survival amplitude itself has a Gaussian distribution.
\\

Using Eq.~(\ref{time_aver-cos-sin}), the time averages of the real and imaginary parts of the correlation function are given by $\overline{C_{\rm r}(t)}=\vert V_{00}\vert^2$ and $\overline{C_{\rm i}(t)}=0$.  With similar considerations, the variances can be evaluated to be
\be\label{var-Cr-Ci}
\Delta C_{\rm r,i}^2 \equiv \Big\langle\overline{\big[C_{\rm r,i}(t)-\overline{C_{\rm r,i}(t)}\big]^2}\Big\rangle_{\rm GOE} \simeq \frac{3N}{8 a_V^2} \, .
\ee
This behaviour is confirmed by numerical simulations at finite temperatures, as shown in Fig.~\ref{Fig14} plotting the standard deviations $\Delta C_{\rm r,i}$ versus the matrix dimension $N$.

\subsection{The finite-temperature regime}

Examples of probability distributions for the fluctuations in a correlation function are depicted in Fig.~\ref{Fig13}(a) and Fig.~\ref{Fig13}(b) for $0 < \beta < \infty$ (i.e., $0<T<\infty$).
\\

At low temperatures, the probability distributions are close to Gaussian distributions centered around the mean values of the real and imaginary parts of the correlation function. Figure~\ref{Fig13}(a) shows the probability distributions in the case $\beta=50$.  This behaviour is similar as in the zero-temperature regime.  The width of the distributions, i.e., the standard deviation of the fluctuations, increases as $\sqrt{N}$ in agreement with the expectation~(\ref{var-Cr-Ci}) in this regime, as seen in Fig.~\ref{Fig14}.
\\

At high temperatures, the probability distribution of the imaginary part remains Gaussian-like, but the probability distribution of the real part deviates from a Gaussian distribution and becomes asymmetric with an exponential-like tail for positive values of $C_{\rm r}$. See Fig.~\ref{Fig13}(b) in the case $\beta=1$.  This effect is more and more pronounced as the temperature increases, i.e., towards the limit $\beta=0$.  As long as $\beta >0$ (i.e., $T<\infty$), the standard deviation of the fluctuations continues to grow as $\sqrt{N}$ with the size $N$ of the matrices, as Fig.~\ref{Fig14} shows.

\subsection{The infinite-temperature regime}

In the limit $\beta=0$ (i.e., $T=\infty$), the correlation function is given by
\be\label{correl-basis-T=infty}
C(t) = \frac{1}{N} \sum_{k,l} \vert V_{kl}\vert^2 \, {\rm e}^{\frac{\rm i}{\hbar} (E_l-E_k)t} = {\rm Re}\, C(t)
\ee
and its imaginary part is equal to zero, ${\rm Im}\, C(t)=0$.  Accordingly, the only non-trivial probability distribution is the one of its real part $C_{\rm r}(t)$.  This distribution has an exponential tail at positive values of $C_{\rm r}$, which can be proved as follows if $N$ is large enough.  For large values of $C_{\rm r}$, the matrix elements $\vert V_{kl}\vert^2$ can be approximated by their mean value~(\ref{var-V}).  In the GOE ensemble, we would thus have that $\vert V_{kl}\vert^2\simeq (1+\delta_{kl})/(2 a_V)$, giving the following approximation for the real part of the correlation function
\be\label{correl-basis-T=infty-approx}
C_{\rm r}(t) = {\rm Re}\, C(t) \simeq \frac{1}{2N a_V} \bigg\vert \sum_{l} {\rm e}^{\frac{\rm i}{\hbar} E_l t} \bigg\vert^2 + \frac{1}{2 a_V} \, .
\ee
Therefore, the real part of the correlation function can be written as
\be\label{correl-basis-T=infty-approx}
C_{\rm r} \simeq \frac{1}{2N a_V} \big( A_{\rm r}^2 + A_{\rm i}^2 \big) + \frac{1}{2 a_V}
\ee
in terms of
\be
A_{\rm r} \equiv \sum_l \cos\left( \frac{E_l}{\hbar}\, t \right)
\qquad\mbox{and}\qquad
A_{\rm i} \equiv \sum_l \sin\left( \frac{E_l}{\hbar}\, t \right) .
\ee
Since we have the time averages $\overline{A_{\rm r}(t)}=\overline{A_{\rm i}(t)}=0$, $\overline{A_{\rm r}(t)^2}=\overline{A_{\rm i}(t)^2}=N/2$, and $\overline{A_{\rm r}(t)A_{\rm i}(t)}=0$, we may consider $A_{\rm r}$ and $A_{\rm i}$ as statistically independent Gaussian random variables with the following distributions,
\be
{\cal P}(A_{\rm r,i}) = \frac{1}{\sqrt{\pi N}} \, \exp\left(-\frac{ A_{\rm r,i}^2}{N}\right) .
\ee
As a consequence, the real part~(\ref{correl-basis-T=infty-approx}) has the following probability distribution at large positive values,
\be\label{exp-tail}
{\cal P}(C_{\rm r}) \simeq 2 a_V \, \exp\left( -2 a_V C_{\rm r} + 1 \right) 
\qquad\mbox{for}\quad
C_{\rm r} \to +\infty \, .
\ee
Since this exponential tail does not depend on $N$, the variance becomes quasi independent of the matrix size $N$ in the infinite-temperature limit.  The predictions for the exponential tail~(\ref{exp-tail}) and the corresponding variance are confirmed by numerical simulations, as observed in Fig.~\ref{Fig13}(c) and Fig.~\ref{Fig14}.
\\

Similar results hold for the GUE and GSE ensembles.
\\

Here, the conclusion is that the quantum fluctuations persist over arbitrarily long time intervals with standard deviations going as $\sqrt{N}$, although the correlation function behaves as $N$ at early and intermediate times.  Accordingly, the quantum fluctuations become negligible as $N\to\infty$ and, for large enough systems, the correlation function of an individual member of the ensemble is well approximated by the ensemble average.

%%%%%%%%%%%%%%%%%%%%%%%%%%%%%%%%%%%%%%%%%%%%%%%%%
\section{Conclusion and perspectives}
\label{sec:conclusion}

In this paper, we have studied the dynamical response and correlation functions of random quantum systems defined in the framework of random-matrix theory and modeling disordered systems such as infinitely many randomly distributed isolated $N$-level subsystems, as explained in Sec.~\ref{sec:systems}.  We have shown in Sec.~\ref{sec:response} that the response functions of these infinite quantum systems can be exactly expressed as the statistical average of the response functions of the individual parts over the random-matrix ensemble.  In Sec.~\ref{sec:1st-resp}, we have investigated the properties of the first-order response function, the two-time correlation function, and the associated spectral density at different temperatures for the orthogonal, unitary, and symplectic Wigner-Dyson universality classes.  The intermediate-time behaviour of the mean correlation function was shown to be given in relation to the semicircle law for the DoS of the Gaussian ensembles considered.  At positive temperatures, we deduced that the mean correlation function has a power-law decay for the GOE, but a much faster decay for the GUE and GSE.  In contrast, at zero temperature, the imaginary part of the mean correlation function has a power-law decay for the GUE and GSE, but a much faster decay for the GOE.  In relation to these universal features at long times, the associated spectral density is shown to have a dip around zero frequency.

In Sec.~\ref{sec:higher-resp}, we have obtained the mean third-order response function and the related mean four-time correlation function that we have explicitly calculated at intermediate times in terms of Bessel functions.  The diagrammatic method has been developed to calculate the multi-time correlation functions at any order.  The enumeration of the diagrams contributing to a given order is related to an interesting combinatoric problem, involving the Bell polynomials \cite{B39,R58}. It is seen that many terms in Bell polynomials are zero. Thus, with the constraint of connecting the vertices with an even number of links, a new set of polynomials have been found. With the help of these polynomials, response functions of any order may be calculated. Further, the calculated response functions are used to obtain the dynamical response of the system to impulsive perturbations in Sec.~\ref{sec:impulsive}.  In Sec.~\ref{sec:fluctuations} devoted to the quantum fluctuations in the individual members of an ensemble, we have found that the probability density of the fluctuations changes as the temperature increases from zero to infinity.  In the latter limit, the probability density has an exponential tail at positive values, which we have quantitatively characterised.

Our results extend to the non-Gaussian random-matrix ensembles considered in Refs.~\cite{BZ93,BZ94}.  Furthermore, we note that similar results hold for other random-matrix ensembles based on the tenfold way by considering not only time reversal, but also charge conjugation.  Accordingly, there exist ten symmetry classes of matrices in correspondence with the real and complex Clifford algebras, instead of the three classes obtained if only time reversal is considered \cite{AZ97,B20}.  Nevertheless, these ten classes fall into the three Wigner-Dyson universality classes with respect to their level statistics at small energy scales, as summarized in Table~\ref{Tab:AZ}.  Therefore, the mean response and correlation functions of models based on such random-matrix ensembles have the same long-time behaviours as those we have here obtained.

\begin{table}[h!]
\begin{tabular}{c @{\hskip 1cm} c @{\hskip 1cm} c @{\hskip 1cm} c  }
\hline\hline
WD class &   AZ classes     &   $T$  &  $\nu$   	 \\
\hline  
OE & AI, BDI, CI & $1$ & $1$ \\
UE & A, AIII, C, D & $0$ & $2$ \\
SE & AII, CII, DIII & $-1$ & $4$ \\
\hline\hline
\end{tabular}
 \caption{Classification of the Altland-Zirnbauer (AZ) symmetry classes into the three Wigner-Dyson (WD) universality classes.  OE, UE, and SE are respectively the orthogonal, unitary, and symplectic ensembles. $T$ denotes the time-reversal parity, which is equal to zero if there is no such a symmetry \cite{AZ97,B20}.  The AZ classes are denoted with Cartan's notation. $\nu$ is the exponent of the level-spacing distribution at zero spacing in the Wigner-Dyson class.}\label{Tab:AZ}
\end{table}

Now, we come back to the issue raised in the introduction.  The universal features of generic quantum systems revealed by Wigner-Dyson random-matrix theory are associated with small energy scales in the range of the mean level spacing $\langle S\rangle_{E_{\rm th}}=\sigma(E_{\rm th})^{-1}$, which is the inverse of the DoS at the mean thermal energy $E_{\rm th}$.  In the time domain, this small energy scale corresponds to the long-time scale, $t\gg \hbar/\langle S\rangle_{E_{\rm th}}=\hbar\sigma(E_{\rm th})$.  Since the mean level density grows exponentially with the number of degrees of freedom, these universal features become unobservable in many-body systems at the macroscale.  Moreover, the universal effects only concern each one of the possible symmetry sectors of a many-body system.  In general, the observed quantities are given by combinations of all the symmetry sectors, which hides even more the tiny universal effects.  For instance, in the two-dimensional lattice of $N$ spins with dipolar interactions \cite{vEG94}, the dimension of the quantum state space is equal to $2^N$ and the discrete translational and rotational symmetries decompose the Hamiltonian into a number of symmetry blocks of the order of $N$.  Accordingly, each symmetry sector has a dimension increasing like $2^N/N$, where Wigner spacing distributions are observed \cite{vEG94}.  However, the response properties of such many-body systems are mainly understood in terms of the DoS and the operators coupling the spins to external fields.  The Wigner-Dyson universal features are thus observable in few-body systems where the quantization of energy into a discrete spectrum can be observed as in nuclei, atoms, molecules, or nanoparticles, but these features become unobservable for macroscopic condensed matter.

%%%%%%%%%%%%%%%%%%%%%%%%%%%%%%%%%%%%%%%%%%%%%%%%%%
%\pagebreak

\section*{Acknowledgements}

The authors thank late Professor G. Nicolis for support and encouragement in the early stage of this work.
They also thank Professor John W. Turner for some interesting and instructive discussions on
some combinatorial aspects, Professor Stuart A. Rice for
constructive criticism on some parts of this work, as well as Professor Daniel Alonso
for discussions regarding non-Gaussian ensembles. During the early stage of this
work (1995-97), S.R.J. was financially supported by the ``Communaut\'e Fran\c caise
de Belgique'' under contract no. ARC-93/98-166 and P.G. by the
National Fund for Scientific Research (F.N.R.S. Belgium).  

%%%%%%%%%%%%%%%%%%%%%%%%%%%%%%%%%%%%%%%%%%%%%%%%
%\pagebreak

\appendix
 
\section{Properties of random matrices}
\label{app:RMT}

In this appendix, we list results that we use in the main text about the statistical properties of random matrices.
\\

\paragraph{Variances of random-matrix elements.} Equation~(\ref{var-V}) for the variances of the matrix elements of the perturbation operator is obtained starting from the probability distribution~(\ref{P(V)}), where the trace can be written as
\be
{\rm tr}\, \hat V^2 = \sum_k V_{0,kk}^2 + 2 \sum_{k<l} \left( V_{0,kl}^2+\cdots + V_{\nu-1,kl}^2\right)
\ee
for orthogonal ($\nu=1$), unitary ($\nu=2$), and symplectic ($\nu=4$) Gaussian ensembles.  Here, we use the notations $V_{0,kl}=V_{kl}$ for the orthogonal ensemble, $V_{0,kl}={\rm Re}\, V_{kl}$ and $V_{1,kl}={\rm Im}\, V_{kl}$ for the unitary ensemble, and those of Eq.~(\ref{symplectic}) with $V_{j,kl}=\langle k \vert \hat V_j\vert l\rangle$ and $j=0,1,2,3$ for the symplectic ensemble.  Accordingly, we find the variances $\langle \vert V_{kk}\vert^2\rangle=\langle V_{0,kk}^2\rangle=(\nu a_V)^{-1}$, if $k=l$; and $\langle \vert V_{kl}\vert^2\rangle=\langle V_{0,kl}^2+\cdots+V_{\nu-1,kl}^2\rangle=(2 a_V)^{-1}$, if $k\ne l$.  These values can be combined with each others into Eq.~(\ref{var-V}).
\\
 
\paragraph{Spacing distributions.} For the three Wigner-Dyson universality classes, the nearest-neighbour spacing probability distributions are surmised by
\begin{eqnarray}
{\cal P}(s) &=& \frac{\pi \, s}{2} \exp \left( -\frac{\pi s^2}{4} \right)
\qquad{\mbox{(orthogonal)}} ,\label{P(s)-OE} \\
{\cal P}(s) &=& \frac{2^5 s^2}{\pi ^2}\exp \left( -\frac{4 s^2}{\pi} \right)
\qquad{\mbox{(unitary)}} , \label{P(s)-UE} \\
{\cal P}(s) &=& \frac{2^{18} s^4}{3^6\pi ^3}\exp \left( -\frac{64 s^2}{9\pi} \right)
\qquad{\mbox{(symplectic)}} , \label{P(s)-SE}
\end{eqnarray}
after a rescaling such that $\langle s\rangle=1$ \cite{H01}.

%%%%%%%%%%%%%%%%%%%%%%%%%%%%%%%%%%%%%%%%%%%%%%%%

\section{Long-time behaviour for the Gaussian ensembles}
\label{app:long-time}

In order to evaluate the mean correlation function~(\ref{Mcorrel-sigmas-approx}) at positive temperatures, we introduce
\be
\Delta E = E-E' 
\qquad\mbox{and}\qquad
{\cal E} = \frac{E+E'}{2} \, ,
\ee
so that
\be\label{De}
\Delta e = e-e'= \frac{\sigma({\cal E})+{\cal E}\, \partial_{\cal E}\sigma({\cal E})}{g_\nu} \, \Delta E + O(\Delta E^3) \, .
\ee
As a consequence, Eq.~(\ref{Mcorrel-sigmas-approx}) becomes
\be\label{Mcorrel-sigmas-approx-2}
\langle C(t)\rangle_{\rm ME} \simeq \frac{\gamma_\nu + g_\nu}{2 a_V} + \frac{1}{2 a_V} \, I(t)
\ee
with the integral
\be\label{I(t)-1}
I(t) = \frac{1}{\langle Z(\beta)\rangle_{\rm ME}} \int_{-E_{\rm b}}^{+E_{\rm b}} d{\cal E} \, {\rm e}^{-\beta{\cal E}}  \int d\Delta E \, \sigma\Big({\cal E}+\frac{\Delta E}{2}\Big)\, \sigma\Big({\cal E}-\frac{\Delta E}{2}\Big) \left[ 1-Y(\Delta e)\right] {\rm e}^{-\beta\frac{\Delta E}{2}} \, {\rm e}^{-\frac{\rm i}{\hbar} \Delta E\, t} \, .
\ee
Here, we may introduce the new integration variable $x=\Delta E\, t/\hbar$, showing that the asymptotic value of the integral is controlled in the long-time limit by its behaviour at small values for $\Delta E =\hbar x/t$, where the two-point function $Y(\Delta e)$ is given by Eq.~(\ref{Y-small}).  Accordingly, the integral is evaluated as
\be\label{I(t)-2}
I(t) = \frac{\alpha_\nu}{\langle Z(\beta)\rangle_{\rm ME}} \left(\frac{\hbar}{t}\right)^{\nu+1}\int_{-E_{\rm b}}^{+E_{\rm b}} d{\cal E} \, {\rm e}^{-\beta{\cal E}}  \int dx \, \sigma\Big({\cal E}+\frac{\hbar x}{2t}\Big)\, \sigma\Big({\cal E}-\frac{\hbar x}{2t}\Big) \left\vert\frac{\sigma({\cal E})+{\cal E}\, \partial_{\cal E}\sigma({\cal E})}{g_\nu}\right\vert^{\nu} \vert x\vert^{\nu}\, {\rm e}^{-\beta\frac{\hbar x}{2t}} \, {\rm e}^{-{\rm i}x} \left[ 1 + O\Big(\frac{x^2}{t^2}\Big)\right]
\ee
with a correction going as $O(\frac{x^2}{t^2})$ because of Eq.~(\ref{De}).  Consequently, the next-to-leading term in the expansion $\exp(-\beta\frac{\hbar x}{2t})=1-\beta\frac{\hbar x}{2 t} + O(\frac{x^2}{t^2})$ gives the only next-to-leading term going as $O(\frac{x}{t})$ in the asymptotic expansion of Eq.~(\ref{I(t)-2}).  We thus obtain
\be\label{I(t)-3}
I(t) = 2 \, A_\nu \left[ \Gamma(\nu+1) \, \cos\frac{\pi(\nu+1)}{2} \left(\frac{\hbar}{t}\right)^{\nu+1} + {\rm i} \, \frac{\beta}{2} \, \Gamma(\nu+2) \, \sin\frac{\pi(\nu+2)}{2} \left(\frac{\hbar}{t}\right)^{\nu+2}+ O\left(\frac{1}{t^{\nu+3}}\right)\right]
\ee
with the following coefficient,
\be\label{coeff-A_nu}
A_\nu = \frac{\alpha_\nu}{\langle Z(\beta)\rangle_{\rm ME}} \int_{-E_{\rm b}}^{+E_{\rm b}} d{\cal E} \, {\rm e}^{-\beta{\cal E}} \sigma({\cal E})^2 \left\vert\frac{\sigma({\cal E})+{\cal E}\, \partial_{\cal E}\sigma({\cal E})}{g_\nu}\right\vert^{\nu} \, ,
\ee
which leads to the result~(\ref{Mcorrel-long-time}).

%%%%%%%%%%%%%%%%%%%%%%%%%%%%%%%%%%%%%%%%%%%%%%%%

\section{Long-time behaviour for the Gaussian ensembles with $N=2$}

Here, we shall use the notation $a= a_{H_0}$ to denote the parameter of the Gaussian distributions~(\ref{P(H0)}).

\subsection{GOE two-level systems}
\label{app:GOE.N=2}

For the Gaussian orthogonal ensemble ($\nu=1$) of $2\times 2$ matrices ($N=2$), the mean spacing is given by $\langle S\rangle_{\rm GOE}=\sqrt{\pi/a}$.  Accordingly, the rescaled spacing distribution~(\ref{P(s)-OE}) leads to
\be
{\cal P}(S) = \frac{a}{2} \, S \, \exp\left(-\frac{a}{4}\, S^2\right)
\ee
for the spacing $S=s \langle S\rangle_{\rm GOE}$ itself.

The average of the cosine function giving the real part of the mean correlation function is thus obtained in terms of the integral
\be
\left\langle\cos\frac{St}{\hbar}\right\rangle_{\rm GOE} = \int_0^{\infty} dS \, {\cal P}(S) \, \cos\frac{St}{\hbar} = \int_0^{\infty} dx \, {\rm e}^{-x} \, \cos\left(\frac{2t}{\hbar}\sqrt{\frac{x}{a}}\right) ,
\ee
using the new integration variable $x=aS^2/4$.  Expanding the cosine function in Taylor series, the result can be expressed with the Kummer function
\be
\left\langle\cos\frac{St}{\hbar}\right\rangle_{\rm GOE} = M\Big( 1,\frac{1}{2}, - \frac{t^2}{a\hbar^2} \Big) \, ,
\ee
which is equal to the confluent hypergeometric function $_1F_1(\alpha;\beta;z)=M(\alpha,\beta,z)$ \cite{AS72,GR80}.
\\

The asymptotic expansion of this function can be deduced noting that $S\cos(St/\hbar)=\hbar\partial_t\sin(St/\hbar)$.  Introducing the new integration variable $x=S\sqrt{a/2}$ and the new time $\tau=\sqrt{2/a}(t/\hbar)$, the average of the cosine function can be expressed as
\be
\left\langle\cos\frac{St}{\hbar}\right\rangle_{\rm GOE} = \partial_\tau \, J(\tau)
\qquad\mbox{with}\qquad
J(\tau) \equiv \int_0^{\infty} dx \, {\rm e}^{-\frac{x^2}{2}} \, \sin x \tau \, .
\label{GOE-cos-J-dfn}
\ee
The asymptotic expansion of the function $J(\tau)$ is calculated here below in Appendix~\ref{app:J(tau)}.  Using the result~(\ref{asympt-expansion-J}), we obtain
\be
\left\langle\cos\frac{St}{\hbar}\right\rangle_{\rm GOE} = -\frac{1}{\tau^2} -\frac{3}{\tau^4} -\frac{15}{\tau^6} -\frac{105}{\tau^8}-\frac{945}{\tau^{10}} - \cdots + O\left(\tau\, {\rm e}^{-\frac{\tau^2}{2}}\right)
\ee
and, therefore,
\be\label{asympt-expansion-f(t)}
\left\langle\cos\frac{St}{\hbar}\right\rangle_{\rm GOE} = -\frac{a\hbar^2}{2t^2} - 3\left(\frac{a\hbar^2}{2t^2}\right)^2 - 15 \left(\frac{a\hbar^2}{2t^2}\right)^3 - 105 \left(\frac{a\hbar^2}{2t^2}\right)^4  - 945 \left(\frac{a\hbar^2}{2t^2}\right)^5  - \cdots + O\left( t \, {\rm e}^{-\frac{t^2}{a\hbar^2}}\right) .
\ee
Keeping the leading term gives the first line of Eq.~(\ref{Re-Im-Mcorrel-GOE-N=2}) with $a=a_{H_0}$ for the long-time behaviour of the real part of the mean correlation function.

Now, inserting the expansion~(\ref{asympt-expansion-f(t)}) into the last formula in the second line of Eq.~(\ref{Re-Im-Mcorrel-levels-N=2}) leads to the second line of Eq.~(\ref{Re-Im-Mcorrel-GOE-N=2}) for the long-time behaviour of the imaginary part of the mean correlation function at positive temperatures.
\\

In the zero-temperature limit, the imaginary part of the mean correlation function is given by the statistical average of the sine function according to
\be
\left\langle\sin\frac{St}{\hbar}\right\rangle_{\rm GOE} = -\frac{a\hbar}{2} \, \frac{\partial}{\partial\tau}\left[ \int_0^{\infty} dS \, \exp\left(-\frac{a}{4} S^2\right) \cos\frac{St}{\hbar} \right] = -\frac{\hbar}{2} \, \sqrt{\pi a} \, \frac{\partial}{\partial\tau} \exp\left(-\frac{\tau^2}{2}\right) ,
\ee
which leads to the result~(\ref{ImC-T=0.GOE}).

\subsection{GUE two-level systems}
\label{app:GUE.N=2}

For the Gaussian unitary ensemble ($\nu=2$) of $2\times 2$ matrices ($N=2$), the mean spacing is given by $\langle S\rangle_{\rm GUE}=\sqrt{8/(a\pi)}$.  According to Eq.~(\ref{P(s)-UE}), the spacing $S=s \langle S\rangle_{\rm GUE}$ has the following distribution,
\be
{\cal P}(S) = a\sqrt{\frac{2a}{\pi}} \, S^2 \, \exp\left(-\frac{a}{2}\, S^2\right) .
\ee
On the one hand, the average of the cosine function can be obtained as
\be
\left\langle\cos\frac{St}{\hbar}\right\rangle_{\rm GUE} = -2a\sqrt{\frac{2a}{\pi}} \, \frac{\partial}{\partial a}\left[ \int_0^{\infty} dS \, \exp\left(-\frac{a}{2}\, S^2\right) \cos\frac{St}{\hbar} \right] ,
\ee
exactly giving
\be
\left\langle\cos\frac{St}{\hbar}\right\rangle_{\rm GUE} = \left(1-\frac{t^2}{a\hbar^2}\right) \, \exp\left(-\frac{t^2}{2a\hbar^2}\right) ,
\ee
and thus the first line of Eq.~(\ref{Re-Im-Mcorrel-GUE-N=2}).  The second line is deduced using the Taylor expansion in powers of~$\beta$ for the imaginary part in Eq.~(\ref{Re-Im-Mcorrel-levels-N=2}) and keeping the leading term going as $\beta\hbar\partial_t$.  The next-to-leading term involving the third derivative with respect to time is a correction of order $O(\beta^2/a)$ relative to the leading term.
\\

On the other hand, the average of the sine function can be expressed as
\be
\left\langle\sin\frac{St}{\hbar}\right\rangle_{\rm GUE} = - a\sqrt{\frac{2a}{\pi}} \, \left(\hbar\frac{\partial}{\partial\tau}\right)^2\left[ \int_0^{\infty} dS \, \exp\left(-\frac{a}{2}\, S^2\right) \sin\frac{St}{\hbar} \right] = - \sqrt{\frac{2}{\pi}} \, \partial_{\tau}^2 \, J(\tau)
\ee
in terms of the new time $\tau = t/(\hbar\sqrt{a})$ and the function $J(\tau)$ introduced in Eq.~(\ref{GOE-cos-J-dfn}) with the new integration variable $x=S\sqrt{a}$.  The asymptotic expansion of the function $J(\tau)$ is obtained in Appendix~\ref{app:J(tau)}, giving
\be\label{GUE-<sin>}
\left\langle\sin\frac{St}{\hbar}\right\rangle_{\rm GUE} = - 2\, \sqrt{\frac{2}{\pi}} \left[ \frac{1}{\tau^3} + \frac{6}{\tau^5} + \frac{45}{\tau^7} + \frac{420}{\tau^9} + \cdots + O\left({\rm e}^{-\frac{\tau^2}{2}}\right)\right] ,
\ee
from which Eq.~(\ref{ImC-T=0.GUE}) is obtained in the zero-temperature limit.

\subsection{GSE two-level systems}
\label{app:GSE.N=2}

This case is similar to the previous case because the spacing distribution here also vanishes as an even power of the spacing. Indeed, for the Gaussian symplectic ensemble ($\nu=4$) of $2\times 2$ matrices ($N=2$), the mean spacing takes the value $\langle S\rangle_{\rm GSE}=8/(3\sqrt{\pi a})$.  According to Eq.~(\ref{P(s)-SE}), the distribution of the spacing $S=s \langle S\rangle_{\rm GSE}$ is here given by
\be
{\cal P}(S) = C \, S^4 \, \exp\left(-a\, S^2\right)
\qquad\mbox{with}\qquad
C=\frac{8 a^2}{3}\sqrt{\frac{a}{\pi}} \,  .
\ee
On the one hand, the average of the cosine function can thus be calculated as
\be
\left\langle\cos\frac{St}{\hbar}\right\rangle_{\rm GSE} = C \, \frac{\partial^2}{\partial a^2}\left[ \int_0^{\infty} dS \, \exp\left(-a\, S^2\right) \cos\frac{St}{\hbar} \right] ,
\ee
leading to the following exact result,
\be
\left\langle\cos\frac{St}{\hbar}\right\rangle_{\rm GSE} = \left(1- \frac{t^2}{a \hbar^2} +\frac{t^4}{12 a^2 \hbar^4}\right) \, \exp\left(-\frac{t^2}{4a\hbar^2}\right)\, ,
\ee
and the first line of Eq.~(\ref{Re-Im-Mcorrel-GSE-N=2}).  As in the other cases, the second line is deduced using the Taylor expansion in powers of~$\beta$ for the imaginary part in Eq.~(\ref{Re-Im-Mcorrel-levels-N=2}) and keeping the leading term going as $\beta\hbar\partial_t$, the next-to-leading term being a correction of relative order $O(\beta^2/a)$.
\\

On the other hand, the average of the sine function is given by
\be
\left\langle\sin\frac{St}{\hbar}\right\rangle_{\rm GSE} = C \, \left(\hbar\frac{\partial}{\partial\tau}\right)^4\left[ \int_0^{\infty} dS \, \exp\left(-a\, S^2\right) \sin\frac{St}{\hbar} \right] = \frac{1}{3}\, \sqrt{\frac{2}{\pi}} \, \partial_{\tau}^4 \, J(\tau)
\ee
with the new time $\tau = t/(\hbar\sqrt{2 a})$ and again the function $J(\tau)$ introduced in Eq.~(\ref{GOE-cos-J-dfn}) but with the new integration variable $x=S\sqrt{2 a}$.  Using the asymptotic expansion of the function $J(\tau)$ given in Appendix~\ref{app:J(tau)}, we obtain
\be\label{GSE-<sin>}
\left\langle\sin\frac{St}{\hbar}\right\rangle_{\rm GSE} = 8\, \sqrt{\frac{2}{\pi}} \left[ \frac{1}{\tau^5} + \frac{15}{\tau^7} + \frac{210}{\tau^9} + \frac{3150}{\tau^{11}} + \cdots + O\left({\rm e}^{-\frac{\tau^2}{2}}\right)\right] ,
\ee
leading to the result~(\ref{ImC-T=0.GSE}) in the zero-temperature limit.
\\

\subsection{Asymptotic expansion of the function $J(\tau)$}
\label{app:J(tau)}

We consider the following function previously introduced in Eq.~(\ref{GOE-cos-J-dfn}),
\be\label{dfn-J}
J(\tau) \equiv \int_0^{\infty} dx \, {\rm e}^{-\frac{x^2}{2}} \, \sin x \tau
\ee
in the limit $\tau\to\infty$.  With an integration by parts, it turns out that this function obeys the following ordinary differential equation,
\be
\frac{d J(\tau)}{d\tau} = -\tau \, J(\tau) - 1 \, .
\ee
Solving this equation gives
\be
J(\tau) = X(\tau) \, \exp\left(-\frac{\tau^2}{2}\right)
\qquad\mbox{with}\qquad
X(\tau)=X(\tau_0)-\int_{\tau_0}^{\tau} \exp\left(\frac{\tau^{\prime 2}}{2}\right) d\tau^{\prime}\, .
\ee
The integral can be expressed by the following asymptotic expansion
\be
\int_{\tau_0}^{\tau} \exp\left(\frac{\tau^{\prime 2}}{2}\right) d\tau^{\prime}
=\exp\left(\frac{\tau^{2}}{2}\right) \sum_{n=1}^{\infty} \frac{b_n}{\tau^n} + C \, ,
\ee
where the coefficients should satisfy the recurrence $b_{n+1} = (n-1) \, b_{n-1}$ starting from $b_1=1$
and $C$ is an integration constant, so that
\be
b_2 = 0 \, , \ b_3 = b_1 = 1 \, , \
b_4 = 0 \, , \ b_5 = 3 \, b_3 = 3 \, , \
b_6 = 0 \, , \ b_7 = 5 \, b_5 = 15 \, , \
b_8 = 0 \, , \ b_9 = 7 \, b_7 = 105 \, , \
\dots
\ee
Therefore, we find that
\be\label{asympt-expansion-J}
J(\tau) = \frac{1}{\tau} + \frac{1}{\tau^3} + \frac{3}{\tau^5} + \frac{15}{\tau^7} + \frac{105}{\tau^9} + \cdots + O\left({\rm e}^{-\frac{\tau^2}{2}}\right) \, .
\ee

%%%%%%%%%%%%%%%%%%%%%%%%%%%%%%%%%%%%%%%%%%%%%%%%

\section{Spectral density near its edges}
\label{app:spctrl_density}

If the frequency is positive ($\omega >0$), the overlap between the two semicircular densities in Eq.~(\ref{spctrl_density_large_freq}) for the mean spectral density extends over the energy interval $-E_{\rm b} + \hbar\omega < E < +E_{\rm b}$.  The range of this interval is equal to $\Delta = 2 E_{\rm b}-\hbar\omega$, which vanishes as $\hbar\omega\to 2 E_{\rm b}$.  Near this edge, the product of the two semicircular densities can be approximated as follows,
\bea
\sigma(E)\, \sigma(E-\hbar\omega) &=& \left(g_\nu \frac{a_{H_0}}{\pi}\right)^2 \, \sqrt{E_{\rm b}^2 - E^2} \, \sqrt{E_{\rm b}^2 - (E-\hbar\omega)^2} \nonumber\\
&=& \left(g_\nu \frac{a_{H_0}}{\pi}\right)^2 \, \sqrt{\left(2E_{\rm b}-\frac{\Delta}{2}\right)^2 - \epsilon^2} \, \sqrt{\left(\frac{\Delta}{2}\right)^2 - \epsilon^2} \nonumber\\
&\simeq& \left(g_\nu \frac{a_{H_0}}{\pi}\right)^2 \, 2 E_{\rm b} \, \sqrt{\left(\frac{\Delta}{2}\right)^2 - \epsilon^2} 
\qquad\mbox{for}\qquad
\vert\Delta\vert \ll E_{\rm b} \, ,
\eea
where $\epsilon=E-E_{\rm b} + \frac{\Delta}{2} \in\{ - \frac{\Delta}{2},+\frac{\Delta}{2}\}$.  Accordingly, the mean spectral density~(\ref{spctrl_density_large_freq}) can be evaluated near its edge as
\be\label{spctrl_density_large_freq-2}
\langle S(\omega)\rangle_{\rm ME} \simeq  \frac{\pi\hbar}{a_V \langle Z(\beta)\rangle_{\rm ME}} \left(g_\nu \frac{a_{H_0}}{\pi}\right)^2 \, 2 E_{\rm b} \, {\rm e}^{-\beta \left( E_{\rm b}-\frac{\Delta}{2}\right)}  \int_{-\frac{\Delta}{2}}^{+\frac{\Delta}{2}} d\epsilon \, {\rm e}^{-\beta \epsilon} \, \sqrt{\left(\frac{\Delta}{2}\right)^2 - \epsilon^2}  \, ,
\ee
which gives
\be\label{spctrl_density_large_freq-3}
\langle S(\omega)\rangle_{\rm ME} \simeq  \frac{\pi\hbar}{a_V \langle Z(\beta)\rangle_{\rm ME}} \left(g_\nu \frac{a_{H_0}}{\pi}\right)^2 \, 2 E_{\rm b} \, {\rm e}^{-\beta E_{\rm b}} \, \frac{\pi}{2}  \left(\frac{\Delta}{2}\right)^2 \, \left[1 + O(\beta\Delta)\right] ,
\ee
hence the result~(\ref{spctrl_density_edges}) near the edge at the positive frequency $\hbar\omega=+2E_{\rm b}$.  A similar calculation leads to the result~(\ref{spctrl_density_edges}) near the other edge at $\hbar\omega=-2E_{\rm b}$.

%%%%%%%%%%%%%%%%%%%%%%%%%%%%%%%%%%%%%%%%%%%%%%%%
%\newpage

%%%%%%%%%%%%%%%%%%%%%%%%%%%%%%%%%%%%%%%%%%%%%%%%%%%%%%%%%%

\end{document}